\documentclass[11pt,a4paper]{article}
\usepackage{jheppub}
\usepackage{xcolor,bm,multirow} 
\usepackage{caption, subcaption} 
 \newcommand{\eir}{\epsilon_{\mathrm{IR}}}

\newcommand{\mms}{\mu^{2\epsilon}_{\overline{\mathrm{MS}}}}
\newcommand{\dms}{\frac{(\mu^2 e^{\gamma_{\mathrm{E}}})^{\epsilon}}{\Gamma(1-\epsilon)}}
\newcommand{\dd}[1]{\frac{d^D {#1} }{(2\pi)^D}} 
\newcommand{\scone}{$\mathrm{SCET_I}$}
\newcommand{\sctwo}{$\mathrm{SCET_{II}}$}

\newcommand{\fms}[1]{{#1}\!\!\!/}

\newcommand{\mc}{\mathcal}

\newcommand{\eps}{\epsilon}  
\begin{document} 
\title{$\bm{N}$-jettiness for muon jet pairs in electroweak high-energy processes}
\def\KU{Department of Physics, Korea University, Seoul 02841, Korea} 
 \def\Seoultech{Institute of Convergence Fundamental Studies and School of Liberal Arts, 
Seoul National University of Science and 
Technology, Seoul 01811, Korea}

\author[a]{Junegone Chay}
\emailAdd{chay@korea.ac.kr}
\affiliation[a]{\KU}
\author[b]{Taehyun Kwon}
\emailAdd{thgwon@seoultech.ac.kr}
\affiliation[b]{\Seoultech}
 
\abstract{We study the $N$-jettiness in the electroweak high-energy process for the final muon jet pairs, 
$e^- e^+ \rightarrow \mu^+ \ \mathrm{jet} + \mu^- \mathrm{jet}$. Compared to QCD, the main difference 
 is that there exist additional gauge nonsinglet contributions in the weak interaction,
which make the factorization more elaborate.  Especially the nonsinglet contributions  arise due to the  
Block-Nordsieck violation in electroweak processes, which yields the Sudakov logarithms and the 
 rapidity divergence. They change the evolution of the factorized parts considerably
in the $N$-jettiness.  There are two possible channels, initiated from the gauge bosons 
$W W \rightarrow \ell_{\mu} \overline{\ell}_{\mu}$, and  from the electrons 
$\ell_e \overline{\ell}_e \rightarrow \ell_{\mu} \overline{\ell}_{\mu}$, where $\ell$ denotes the
weak doublet. The latter was discussed previously, 
and we complete the analysis by studying the first. 
The factorization  for  $W W \rightarrow \ell_{\mu} \overline{\ell}_{\mu}$ 
can be proceeded in a similar way as in the factorization for 
$\ell_e\overline{\ell}_e \rightarrow \ell_{\mu} \overline{\ell}_{\mu}$,  and the result exhibits 
a rich structure.
The new ingredients in this study consist of the $W$ beam functions, and the 
complex color structure of the soft functions and the hard functions.
The resummation of the large logarithms is performed by solving the renormalization group equations
with respect to the renormalization scale and the rapidity scale. In the numerical analysis, we confine 
to the SU(2) weak gauge interaction, and the numerical results are presented for both channels at 
next-to-leading-logarithmic accuracy including the singlet and the nonsinglet contributions. The nonsinglet
contributions turn out to be appreciable in the 2-jettiness.}


\maketitle
  
\section{Introduction\label{intro}} 

In high-energy collisions, energetic particles are produced as collimated beams of particles, in the form of jets.
The interplay between the constituent partons participating in the hard scattering and 
the resultant hadrons forming jets after hadronization is intertwined by the perturbative and 
the nonperturbative effects of quantum chromodynamics (QCD).  Since the strong interaction is involved 
at every stage in the scattering process with disparate energy scales, it is difficult to  disentangle 
the interlocking aspects of the strong interaction. 

The proof of the factorization theorems in which the hard, collinear and soft parts are separated in
high-energy scattering has been one of the biggest challenges in QCD.  
In a factorized process, the hard part describes the contribution from the large energy of order $Q$. 
The collinear part takes care of the energetic, collinear particles, which include incoming beams and 
final-state jets. The soft part depicts the soft emissions interspersed between collinear  directions.   

Soft-collinear effective theory 
(SCET)~\cite{Bauer:2000ew,Bauer:2000yr,Bauer:2001ct,Bauer:2001yt} 
has set a new stage in proving the factorization for various jet observables in collider physics. 
SCET is an effective theory, in which collinear, soft modes are  selected as the relevant degrees of freedom. 
The hard degrees of freedom are integrated out to yield the hard function. The  collinear modes in different 
lightcone directions are decoupled, and by redefining the collinear fields with the soft Wilson lines, the 
soft sector is also decoupled at the Lagrangian level. As a result, the factorization  can be established
more transparently in SCET than in full QCD.
 
In SCET, the phase space is divided into the collinear and the soft regions. The invariant masses of their 
modes may be the same (\sctwo) or different (\scone) depending on the physics we are interested in.
One of the important 
objectives in partitioning the phase space is to describe the physics with a single scale in each kinematic 
region, and perform the resummation of the large logarithms by solving the renormalization (RG) equation.  

We step up a gear to consider what happens in electroweak processes at extremely high-energy of order  
10 TeV. Still, the effect of the strong interaction is obviously dominant at LHC. But some electroweak process 
can be observed  in high-energy electron-positron colliders such as 
CEPC~\cite{dEnterria:2016sca},  ILC~\cite{Djouadi:2007ik}, FCC-ee~\cite{Abada:2019zxq}, and
CLIC~\cite{Charles:2018vfv}. 
As an example, the process $e^- e^+ \rightarrow \mu^- \mu^+$ can be described in analogy with the
QCD process $q\overline{q}\rightarrow q' \overline{q}'$.  The incoming ``partons'' in an electron and a positron
possess certain energy fractions, and they emit gauge bosons to be far off-shell before they participate in the 
hard scattering. 
This process is described by the weak beam functions. The energetic partons undergo a hard scattering 
and the final-state particles are observed in terms of  jets, described  by the jet functions.  And the soft function
describes the emission of soft particles  between the collinear particles. 

We may suspect that it is just another copy of QCD by merely replacing the gauge group, and nothing
in particular can be gained. We claim that it is not true. The main difference lies in the fact that hadrons 
 appear only as color singlets in QCD, but weak doublets  such as the electrons
or the neutrinos can be observed.
Because only the color singlets contribute to the observables in QCD, the beam functions and 
the jet functions are obtained by taking the matrix elements of the relevant operators between the hadronic 
states (color singlets) or between the vacuum. However, both singlet 
and nonsinglet contributions are involved in all these factorized components in electroweak processes, 
and it makes the analysis  more intriguing.
We can cast this issue on the imaginary QCD, in which there is no confinement with free quarks and gluons.
Then color nonsinglet contributions affect the jet observables and we can ask whether the factorization 
still works in a consistent fashion in the presence of these extra contributions. Of course, real QCD does not 
work that way, but we can ask the same question toward the weak interaction.
     
There are nonsinglet contributions in the weak interaction to the beam, jet, and soft functions 
as well as singlet contributions.   More importantly, 
because the initial or final states can have fixed charges, Sudakov logarithms can arise  from the 
non-cancellation of the electroweak logarithms, which is known as the  Block-Nordsieck violation in 
electroweak processes~\cite{Ciafaloni:2000df,
Ciafaloni:2001vt,Ciafaloni:2006qu,Manohar:2014vxa}. Note that the Sudakov logarithms 
from virtual and real contributions in QCD cancel in inclusive processes. Due to this cancellation, it yields the  
Dokshitzer-Gribov-Lipatov-Altarelli-Parisi (DGLAP) evolution of the parton distribution functions (PDF) 
in QCD~\cite{Lipatov:1974qm, Gribov:1972ri, Altarelli:1977zs, Dokshitzer:1977sg}. 
The non-cancellation of the Sudakov logarithms in weak interaction affects the renormalization behavior.
 For example, the nonsinglet PDF satisfies nontrivial RG equations due to the Sudakov 
logarithm, while the singlet PDF still satisfies the DGLAP equation. 
If we observe more exclusive, or differential quantities with kinematic constraints on the real emissions, 
there is only a partial cancellation between the virtual and real contributions, resulting in additional 
Sudakov logarithms. Exclusive jet cross sections, or jet shape observables such as $N$-jettiness are such 
examples. Then there appear Sudakov logarithms even in the singlet case.

Furthermore, there is the issue of rapidity divergence~\cite{Chiu:2011qc}, which arises in SCET due to 
the dissection 
of the phase space. If we add the contributions of the factorized parts from all the phase spaces, the 
rapidity divergence cancels. It is why there is no rapidity divergence in QCD because
the phase space is not divided. But aside from the fact that there is 
no rapidity divergence in the singlet contributions, the rapidity divergence  in the electroweak 
nonsinglet contribution survives. It was first pointed out in ref.~\cite{Manohar:2018kfx} 
that the electroweak nonsinglet PDF has the rapidity divergence. 
Because of the rapidity divergence, the double RG evolutions
with respect to the rapidity scale, as well as the conventional renormalization scale should be solved for
both the collinear functions and the soft functions.
 
In this paper, we consider the $N$-jettiness in electroweak processes, in which the Sudakov logarithms appear
both in the singlet and the nonsinglet contributions, while the rapidity logarithm appears in the nonsinglet 
contributions. These result in part from the Block-Nordsieck violation, and in part from the kinematic 
constraint to extract the $N$-jettiness. 
The $N$-jettiness in electroweak processes at extreme high energies is analyzed for the scattering process
in which the final dijets involve a muon and an anti-muon respectively. In order to address the main differences
between QCD and the weak interaction, we consider the $SU(2)$ weak interaction only.  The extension
to the Standard Model is an interesting issue, but we will not pursue it here. It is enough to consider
SU(2) in considering the features from the nonsinglet contributions. 
And we select the 2-jettiness because it involves four lightlike directions, 
in which the dependence of the various lightcone directions appears in the hard and soft functions. 
It offers a nontrivial check to see the independence of the renormalization  and the rapidity scales
in the 2-jettiness. 

Before we proceed, we need to point out that the electron here means the electron with a cloud of 
gauge bosons and all the possible leptons. It is analogous to the proton in QCD in 
the sense that the proton
contains gluons and all the possible quarks (and antiquarks). The confusion may arise because there is no
distinctive terminology to distinguish the electron cloud, and the electron itself, while a quark describes
a parton in the proton in QCD. From now on, the electron $e$ corresponds to the proton in QCD, while
the electron doublet $\ell_e$ describes the partons. 

With this terminology in mind, we consider the 2-jettiness in the process $e^- e^+ \rightarrow \mu^-$-jet
+ $\mu^+$-jet. The final states consist of the dijets in which there is at least a muon and antimuon
in each jet. In our previous paper~\cite{Chay:2021arz}, the factorization for the channel 
$\ell_e \overline{\ell}_e \rightarrow \ell_{\mu} \overline{\ell}_{\mu}$ has been analyzed to 
show the characteristics of  the nonsinglet nature in the weak interaction. In order to analyze the 
2-jettiness for the $\mu^- \mu^+$ jets, the remaining channel 
$W W \rightarrow \ell_{\mu}\overline{\ell}_{\mu}$ 
should be included, which is the main issue of the paper. 
In the factorization for this channel, we need the additional input of  the gauge-boson 
beam functions, and the hard and soft functions associated with this channel.
Especially the color structure in the soft and hard functions is more involved, which
can be expressed in terms of the $3\times 3$ matrices representing the space of the operators in the channel.
After performing the computation at next-to-leading order (NLO), 
we estimate the contributions from the singlets and the nonsinglets with the resummed results 
at next-to-leading logarithmic accuracy (NLL) for these two channels.
 In contrast to the gluon PDF in QCD, the $W$ PDF may be suppressed compared to the 
electron PDF because the $W$ 
bosons are emitted perturbatively from the electron. But we keep the possibility of the appreciable $W$ PDF open
and proceed. In this sense,  the processes $e^+ e^- \rightarrow WW$, $\ell_L W \rightarrow \ell_L W$ 
or $WW \rightarrow WW$, where a muon fragments from the $W$ boson in the final state, are regarded as
the processes at higher orders. 
  
The structure of the paper is as follows: In section~\ref{wwop}, we construct the effective operators 
relevant to the channel $W W\rightarrow \ell_{\mu} \overline{\ell}_{\mu}$ in SCET. 
We choose a basis of operators
proportional to $\delta^{ab}$, $if^{abc}$ and $d^{abc}$, where $f^{abc}$ and $d^{abc}$ are the 
structure constants of the SU($N$) group. This choice of the basis is significant because the symmetric structure 
constants $d^{abc}$ vanish in the SU(2) weak interaction, and it is separated from the 
beginning. In section~\ref{factor}, the factorization theorem for  
$WW \rightarrow \ell_{\mu} \overline{\ell}_{\mu}$ is established. 
By including the previous result of the factorization for 
$\ell_e \overline{\ell}_e \rightarrow \ell_{\mu} \overline{\ell}_{\mu}$, the whole factorization   
for the muon-pair dijets is completed. 
We briefly review  the rapidity divergence in SCET, and discuss how to regulate 
the rapidity divergence in section~\ref{rapdi}. Each factorized part is computed at NLO. 
In section~\ref{bfpdf}, the beam functions and the PDFs for the gauge bosons with the   
matching coefficients are presented.  For completeness,
the muon semi-inclusive jet functions are quoted from ref.~\cite{Chay:2021arz}. 
In section~\ref{hardf}, the hard functions for  $W W \rightarrow \ell_{\mu}\overline{\ell}_{\mu}$ 
are presented,  based on the result from QCD~\cite{Kelley:2010fn}, and the soft function for this process is given 
in section~\ref{softf}. In section~\ref{anomd}, all the anomalous
dimensions are collected, and the RG evolution of the $N$-jettiness is discussed. 
In section~\ref{numeric}, we present a numerical analysis of the 
2-jettiness near threshold for the SU(2) gauge interaction. In section~\ref{conc}, we give a conclusion 
and an outlook.
In appendix~\ref{nlobf}, the detailed computation of the beam functions for the gauge bosons and the PDFs 
at NLO is explained. In appendix~\ref{softzm}, the tree-level soft matrices are listed.

\section{Effective operators for $WW \rightarrow \ell_{\mu}\overline{\ell}_{\mu}$\label{wwop}}
 The effective operators responsible for  $WW \rightarrow \ell_{\mu}\overline{\ell}_{\mu}$  are given as
\begin{equation} \label{oper}
O_I^{\mu\alpha\beta} = \overline{\ell}_{3L} T_I^{ab} \gamma^{\mu} \ell_{4L} 
\mathcal{B}_{1\perp}^{a\alpha} \mathcal{B}_{2\perp}^{b\beta},
\end{equation}
where the collinear-gauge invariant lepton fields $\ell_n$ and the gauge  fields 
$\mathcal{B}_{\perp}^{\mu}$ are defined as
\begin{equation}
\ell_n (x) = W_n^{\dagger} (x) \xi_n (x),  \ \ \mathcal{B}_{n\perp}^{\mu} (x) =\frac{1}{g}
[ W_n^{\dagger} (x) iD_{n\perp}^{\mu} W_n (x)].
\end{equation}
Here $iD_{n\perp}^{\mu} = \mathcal{P}_{n\perp}^{\mu} + gA_{n\perp}^{\mu}$ is the covariant 
derivative.  The collinear Wilson line is given as
\begin{equation} \label{cowil}
W_n (x)= \mathrm{P} \exp \Bigl( ig \int_{-\infty}^0 ds \overline{n} \cdot A_{n } (x+s\overline{n} )
\Bigr)  
=   \sum_{\mathrm{perm.}} \exp\Bigl[ -g \frac{\overline{n} \cdot 
A_{n }(x)}{\overline{n} \cdot \mathcal{P}} \Bigr],
\end{equation}
where P denotes the path ordering along the integration path.

There are three independent operators, and we choose the basis $T_I^{ab}$ $(I=1, 2, 3)$ as
\begin{equation}\label{basis}
T_1^{ab} = \delta^{ab}, \ \ T_2^{ab} = if^{abc} t^c, \ \ T_3^{ab} = d^{abc}t^c,
\end{equation}
where $t^a$ are the SU($N$) generators. Another equivalent  basis for $N\geq 3$ 
can be selected as $\delta^{ab}$, $t^a t^b$ and $t^b t^a$. However, in SU(2) weak interaction, the bases
$\delta^{ab}$, $t^a t^b$ and $t^b t^a$ are no longer independent. In other words, 
$d^{abc}=0$ in SU(2), and there are only two independent operators.  In this case, we  disregard $T_3^{ab}$. 

The soft interactions are decoupled from the collinear fields by the field 
redefinition~\cite{Bauer:2001yt}
\begin{equation} \label{redef}
\ell_{n }^{(0)} (x) = Y_n^{\dagger} (x) \ell_n (x), \ \  
\mathcal{B}_{n\perp}^{\mu (0)} (x) = Y_n^{\dagger} (x)  \mathcal{B}_{n\perp}^{\mu} (x) Y_n (x),
\end{equation}
where the soft Wilson line $Y_n (x)$ in the fundamental representation is given as
\begin{equation} \label{swilfun}
Y_n (x)= \mathrm{P} \exp \Bigl( ig \int_{-\infty}^0 ds n \cdot A_{us}^c t^c (x+sn )
\Bigr) 
=  \sum_{\mathrm{perm.}} \exp \Bigl[ -g \frac{n \cdot A_{us} (x)}{n\cdot \mathcal{P}}   \Bigr].
\end{equation}
For $\mathcal{B}_{n\perp}^{\mu a}$, we can employ the adjoint representation $\mathcal{Y}$ for the
soft Wilson line~\cite{Bauer:2001yt} as
\begin{equation}
\mathcal{B}_{n\perp}^{\mu a (0)} = \mathcal{Y}_n^{ab} \mathcal{B}_{n\perp}^{\mu b},
\end{equation}
where the soft Wilson line $\mathcal{Y}$ is obtained by replacing $t^c$ by the generators $\mathcal{T}^c$
in the adjoint representation with $(\mathcal{T}^c)^{ab}=-if^{cab}$. And the relation
between $\mathcal{Y}$ and $Y$ is given as
$\mathcal{Y}_n^{ab} =\mathrm{tr} [Y_n^{\dagger} t^b Y_n t^a]$. From now on, we use the fields 
after the decoupling and we drop the superscript (0) for simplicity. 

With the field redefinition, the operators in eq.~\eqref{oper} are written as
\begin{equation} \label{opbasis}
O_{I\mu}^{\alpha\beta} = \overline{\ell}_{3L} Y_3^{\dagger} \gamma_{\mu} T_I^{ab} Y_4 \ell_{4L} 
\mathcal{Y}_1^{ac} \mathcal{B}_{1\perp}^{c\alpha} \mathcal{Y}_2^{bd} \mathcal{B}_{2\perp}^{d\beta}.
\end{equation}
In terms of these operators, we can write the effective Lagrangian as
\begin{equation} \label{effl}
\mathcal{L}_{\mathrm{eff}} = -i \sum_I D^{\mu}_{I\alpha\beta} O^{\alpha\beta}_{I\mu}+
\textrm{hermitian\ conjugate},
\end{equation}
where $D^{\mu}_{I\alpha\beta}$ are the Wilson coefficients, which are obtained by integrating out the 
degrees of freedom of order $Q$. 

We consider the $N$-jettiness, which is defined 
as~\cite{Stewart:2010tn, Jouttenus:2011wh} 
\begin{equation} \label{defjet}
\mathcal{T}_N = \sum_k \mathrm{min} \ \Bigl\{ \frac{2 q_i \cdot p_k}{\omega_i} \Bigr\},
\end{equation} 
where $i$ runs over 1, 2 for the beams, and $3, \cdots, N+2$ for the final-state jets.  
Here $q_i$ are the reference momenta of the beams and the jets with the normalization 
factors $\omega_i = \overline{n}_i \cdot q_i$, and $p_k$ are the momenta of all the measured
particles in the final state. 
\begin{equation} \label{omegai}
q_{1,2}^{\mu} =\frac{1}{2} z_{1,2} E_{\mathrm{cm}} n_{1,2}^{\mu}= 
\frac{1}{2} \omega_{1,2} n_{1,2}^{\mu}, \ q_{i}^{\mu} =\frac{1}{2}\omega_i
n_i^{\mu}, (i=3,\cdots, N+2).
\end{equation} 
Here $z_{1,2}$ are the momentum fractions of the beams.
The lightcone vectors $n_1$ and $n_2$  for the beams are  aligned to the $z$ direction, 
$n_1^{\mu} = (1,0,0,1),  \ n_2^{\mu} = (1, 0, 0, -1)$,
and $n_i$ are the lightcone vectors specifying the jet directions. 
The  2-jettiness from the channel $WW\rightarrow \ell_{\mu} \overline{\ell}_{\mu}$ can be 
written as~\cite{Bauer:2008jx}
\begin{equation} \label{observ}
\Bigl(\frac{d\sigma}{d\mathcal{T}_2}\Big)_W = \frac{1}{2s} \int d^4 x \sum_X \langle I| 
\mathcal{L}_{\mathrm{eff}} (x)|
 X \rangle \langle X|\mathcal{L}_{\mathrm{eff}} (0)|I\rangle  
 \delta \Bigl( \mathcal{T}_2 - g (I, X)\Bigr),
\end{equation}
where $|I\rangle$ represents the 
initial state,  $|X\rangle$ denotes the final state, and the  sum over $X$ includes the phase 
space.  The function $g  (I, X)$ extracts the jettiness from the states $|I\rangle$ and $|X\rangle$.  
In SCET, the final states $|X\rangle$ consist of the $n_i$-collinear states $|X_i\rangle$  
and the soft states $|X_s\rangle$.  Since the $n_i$-collinear particles
do not interact with each other, and the soft particles are decoupled from the collinear sectors, 
the final states $|X\rangle$ in the Hilbert space consist of the tensor 
product of the collinear states $|X_i\rangle$ and the soft states $|X_s\rangle$ as
\begin{equation}
|X\rangle = |X_1\rangle \otimes |X_2\rangle \otimes |X_3\rangle \otimes |X_4\rangle \otimes 
|X_s\rangle.
\end{equation}

In order to obtain the 2-jettiness, we multiply $O_{I\mu}^{\alpha\beta}$ and the hermitian conjugate
$O_{J\mu'}^{\dagger\alpha'\beta'}$ with their respective Wilson coefficients and implement the 
measurement of the 2-jettiness. Since the result looks quite complicated, we first consider the product of 
the operators, for simplicity.  It is written as
\begin{align} \label{product}
& \sum_{IJ} D_{J\alpha' \beta'}^{\mu' *} \Bigl( O_{J\mu'}^{\alpha' \beta'} (x) \Bigr)^{\dagger}
D_{I\alpha\beta}^{\mu} O_{I\mu}^{\alpha\beta} (0) \nonumber \\
&= \sum_{IJ} D_{J\alpha' \beta'}^{\mu' *} D_{I\alpha\beta}^{\mu}    \Bigl[  (\overline{\ell}_4 )_{j_4}^{\mu_4} \Bigl( Y_3^{\dagger} T_J^{a'b'} Y_3\Bigr)_{j_4 j_3}
(\gamma_{\mu'})_{\mu_4 \mu_3} (\ell_3)_{j_3}^{\mu_3} \mathcal{Y}_1^{c'a'} \mathcal{B}_{1\perp}^{c'\alpha'}
\mathcal{Y}_2^{b'd'} \mathcal{B}_{2\perp}^{d'\beta'}\Bigr] (x) \nonumber \\
&\times \Bigl[(\overline{\ell}_3)_{i_3}^{\nu_3} \Bigl( Y_3^{\dagger} T_I^{ab} Y_4\Bigr)_{i_3 i_4} 
(\gamma_{\mu})_{\nu_3 \nu_4} (\ell_4)_{i_4}^{\nu_4} \mathcal{Y}_1^{ca} \mathcal{B}_{1\perp}^{c\alpha}
\mathcal{Y}_2^{bd} \mathcal{B}_{2\perp}^{d\beta} \Bigr] (0),
\end{align}
where $i_k$, $j_k$ are the gauge indices, while $\mu_k$, $\nu_k$ are the Dirac
indices. The measurement of the 2-jettiness, $g(I,X)$, 
will be implemented after simplifying this expression.

We rearrange the lepton bilinears in each collinear direction as~\cite{Manohar:2018kfx, Chay:2021arz} 
\begin{align}
\Bigl(\overline{\ell}_n (x)\Bigr)_{\alpha}^i \Bigl( \ell_n (y)\Bigr)_{\beta}^j 
&= (\mathcal{P}_L \fms{n})_{\beta\alpha} \Bigl[ \frac{1}{2N} \delta^{ij} 
\overline{\ell}_n (x) \frac{\fms{\overline{n}}}{2} \ell_n (y) +(t^c)^{ij} \overline{\ell}_n (x) 
\frac{\fms{\overline{n}}}{2} t^c \ell_n (y)\Bigr] \nonumber \\
&= (\mathcal{P}_L \fms{n})_{\beta\alpha}  \sum_c k_{\ell}^c (T^c)^{ij} C_{\ell}^c (x,y),
\end{align}
where $\mathcal{P}_L = (1-\gamma_5)/2$ and we express the generators $(T^0)^{ij} = \delta^{ij}$ 
for the singlet and $(T^c)^{ij} = (t^c)^{ij}$ for the nonsinglets with the corresponding collinear operators
\begin{equation}
C_{\ell}^a (x,y) = \overline{\ell}_n (x) \frac{\fms{\overline{n}}}{2} T^a \ell_n (y).
\end{equation}
The factors $k^a_{\ell}$ are defined as $k^0_{\ell} = 1/(2N)$ for the singlet and 
$k^a_{\ell} = 1$ for the nonsinglets.

The product of the collinear gauge fields can be written as
\begin{align}
\mathcal{B}_{n\perp}^{a\alpha} (x) \mathcal{B}_{n\perp}^{b\beta} (y) 
&= \frac{g_{\perp}^{\alpha\beta}}{D-2}
\Bigl[ \frac{\delta^{ab}}{N^2-1} \Bigl(\delta^{cd}\mathcal{B}_{n\perp}^{c\mu}(x)
\mathcal{B}_{n\perp\mu}^d (y) \Bigr) + 
\frac{Nd^{abc}}{N^2 -4} \Bigl( d^{dec} \mathcal{B}_{n\perp}^{d\mu} (x) \mathcal{B}_{n\perp\mu}^e (y)
\Bigr)  \Bigr] \nonumber \\
& \equiv g_{\perp}^{\alpha\beta} \sum_{m=0}^{N^2-1}   k_W^m (G^m)^{ab}  C_{W}^m (x,y),
\end{align}
where $D$ is the number of the spacetime dimensions.  
The factors $k_W^m$ are given as
\begin{equation}
k_W^m = \left\{ \begin{array}{ll}
\displaystyle \frac{1}{D-2} \frac{1}{N^2 -1}, & m=0, \ (\textrm{singlet}), \\
\displaystyle \frac{1}{D-2} \frac{N}{N^2 -4}, & \ m=1, \cdots, N^2 -1, \ (\textrm{nonsinglets}),
\end{array}
\right.
\end{equation}
and $(G^0)^{ab} = \delta^{ab}$ for the singlet and $(G^m)^{ab} =d^{abm}$ for the nonsinglets.
Note that $d^{abm} =0$ in SU(2), hence there are no nonsinglet beam functions for 
the gauge bosons.  However, we keep these contributions
for the general SU($N$) gauge theory, and put $N=2$ later in the numerical analysis.
And $C_W^m (x,y)$ are the operators, which are defined as 
\begin{equation}
C_W^m (x,y) = (G^m)^{de} \mathcal{B}_{n\perp}^{d\mu} (x) B_{n\perp\mu}^e (y). 
\end{equation}

Using these relations, eq.~\eqref{product} can be written as
\begin{align} \label{squared}
&\sum_{IJ} D_{J\alpha' \beta'}^{\mu' *} D_{I\alpha\beta}^{\mu} \mathrm{tr} (\mathcal{P}_L \fms{n}_4
\gamma_{\mu'}  \fms{n}_3 \gamma_{\mu})  g_{\perp}^{\alpha\alpha'}
g_{\perp}^{\beta\beta'}  \sum_{efgh} k_W^e k_W^f k_{\ell}^g k_{\ell}^h
C_{B_1}^e (x,0) C_{B_2}^f (0, x) C_{\ell_3}^g (0,x) C_{\ell_4}^h (x,0)  \nonumber \\
&\times \Bigl(\mathcal{Y}_1^{c'a'} \mathcal{Y}_2^{b'd'} (x)\Bigr)  (G^e)^{c'c} (G^f)^{dd'}
\Bigl(\mathcal{Y}_1^{ca} \mathcal{Y}_2^{bd}(0) \Bigr) 
 \mathrm{Tr} \, \Bigl[ \Bigl(T^l Y_4^{\dagger} T_J^{a'b'} Y_3(x)\Bigr)   \Bigl(T^m Y_3^{\dagger}
T_I^{ab} Y_4(0)\Bigr)\Bigr].   
\end{align}
This is the matrix element squared to be employed in expressing the $N$-jettiness, which is obtained by 
implementing the measurement of the $N$-jettiness $g(I,X)$ in eq.~\eqref{observ}. 
In order to establish the factorization,
we define the beam functions and the jet functions as the matrix elements of the collinear fields.   
The soft function consists of the soft Wilson lines, and the hard
function is the combination of the Wilson coefficients of the operators, contracted with the appropriate 
Dirac structure.

\section{Factorization of the muon dijet process \label{factor}}
 
Though eq.~\eqref{squared} looks complicated, it can be rearranged to make 
the factorization look manifest. The hard function $H_{IJ}$ is obtained by combining the matching coefficients
$D_{I\alpha\beta}^{\mu}$, contracted with the Dirac structure, and is written as
\begin{equation} \label{hardf}
H_{IJ} =  D_{I\alpha\beta}^{\mu} D_{J\alpha' \beta'}^{\mu' *} g_{\perp}^{\alpha\alpha'}
g_{\perp}^{\beta\beta'}\mathrm{tr} (\mathcal{P}_L \fms{n}_4
\gamma_{\mu'}  \fms{n}_3 \gamma_{\mu}).
\end{equation}
 
Here we follow \scone, in which the hierarchy of the scales is given by $\mathcal{T}^2 \sim M^2 \ll p_c^2
\sim Q\mathcal{T} \ll Q^2$, where $M$ is the mass of the gauge boson. The $n$-collinear momentum scales as $p_n^{\mu} = (\overline{n}\cdot p,
p_{\perp}, n\cdot p)\sim (Q, \sqrt{Q\mathcal{T}}, \mathcal{T})$, while the usoft momentum scales as
$p_{us}^{\mu} \sim (\mathcal{T}, \mathcal{T}, \mathcal{T})$. The $N$-jettiness probes the scale of order
$n\cdot p_n \sim n\cdot p_{us} \sim \mathcal{T}$.

The collinear functions are obtained by taking the matrix elements of the operators $C_W^a$ between 
the gauge bosons to yield the beam functions, and $C_{\ell}^a$ 
between the vacuum to yield the jet function after inserting the measurement of the jettiness.
The beam functions $B_W^a (t, z, \mu, \nu)$ are defined as 
\begin{equation} \label{beam2}
B_W^a (t, z=\omega/P^-, \mu, \nu) = -\omega\langle W| B_{n\perp \mu}^c \delta (t + \omega n \cdot \mathcal{P})
[ \delta (\omega -\overline{n}\cdot \mathcal{P}) B_{n\perp}^{d\mu}] (G^a)^{cd} |W\rangle,
\end{equation}
where $P^- =\overline{n}\cdot P$, and $t=\omega\mathcal{T}$. At tree level, the beam functions are 
normalized as $B_W^{a(0)} (t,z,\mu,\nu) =\delta (t) \delta (1-z)    (G^a)$.

As introduced in ref.~\cite{Chay:2021arz}, we  define the semi-inclusive jet function as
\begin{align} \label{semijet}
J_{l}^a (p^2,  \mu, \nu)  &= \int \frac{d\overline{n}\cdot p_l}{\overline{n} \cdot p_l}
\int \frac{d^2 \mathbf{p}_l^{\perp}}{\omega} \sum_X \mathrm{Tr} \langle 0| 
 \ell_{L} (0) \delta (n\cdot p +n\cdot \mathcal{P}) 
\delta^{(2)}  (p_{\perp} + \mathcal{P}_{\perp})  |l X\rangle\nonumber \\
&\times   \langle l X|[\delta (\omega +\overline{n}\cdot \mathcal{P})
\overline{\ell}_{L} (0) ] \frac{\fms{\overline{n}}}{2}T^a   |0\rangle,
\end{align}
where the lepton $l$ (muon or antimuon) is specified in the final state.
It is normalized at tree level as
$J_l^{a (0)} (p^2) = \delta (p^2) \mathrm{Tr} (T^a P_l)$,
where $P_l$ is the projection operator to the given lepton $l$. 
For example, the projection operator for the muon is given by $P_{\mu} = (1-\sigma^3)/2$ 
in  SU(2), and $P_{\nu_{\mu}} = (1+\sigma^3)/2$
for the muon neutrino.   

The soft functions are given by the vacuum expectation values of the soft part in eq.~\eqref{squared},  
and are given by
\begin{align} \label{softdef}
S_{JI}^{efgh} (\mathcal{T}_s, \mu, \nu ) &= \langle 0| \mathrm{Tr}\, \Bigl[ 
\Bigl(T^g Y_4^{\dagger} T_J^{a'b'} Y_3 \mathcal{Y}_1^{c'a'} \mathcal{Y}_2^{b'd'} (G^e)^{c'c}\Bigr) (x) 
\nonumber\\
&\times \delta \Bigl(\mathcal{T}_s -\sum_{X_s} \mathrm{min} (\{ n_i \cdot p_{X_s} \} )\Bigr) 
\Bigl((G^f)^{dd'} \mathcal{Y}_1^{ca} \mathcal{Y}_2^{bd} T^h Y_3^{\dagger}
T_I^{ab} Y_4 \Bigr) (0)\Bigr].
\end{align}
 
The detailed procedure of the factorization is delineated in detail in ref.~\cite{Chay:2021arz}. Here we 
present the result following the same procedure in the scheme of \scone. The factorization of the 2-jettiness from  
$W W \rightarrow \ell_{\mu} \overline{\ell}_{\mu}$ is written as 
\begin{align} \label{facww}
\Bigl(\frac{d\sigma}{d\mathcal{T}_2}\Bigr)_W &= \frac{1}{2s} \int \frac{dz_1}{z_1} \int\frac{dz_2}{z_2} \int dt_1 dt_2
\int \frac{d^4 p_3}{(2\pi)^4} \frac{d^4 p_4}{(2\pi)^4}(2\pi)^4 \delta^{(4)} \Bigl( \frac{\omega_1 n_1}{2} +\frac{\omega_2 n_2}{2} - \frac{\omega_3 n_3}{2}
-\frac{\omega_4 n_4}{2} \Bigr)  \nonumber \\
&\times \int d\mathcal{T}_s \delta \Bigl( \mathcal{T}_2 - \frac{t_1}{\omega_1} - 
\frac{t_2}{\omega_2} - \frac{p_3^2}{\omega_3} - \frac{p_4^2}{\omega_4} -\mathcal{T}_s\Bigr) 
\sum_{IJ} H_{IJ} S_{JI}^{efgh} (\mathcal{T}_s , \mu, \nu)
\sum_{efgh} k_W^e k_W^f k_{\ell}^g k_{\ell}^h  \nonumber \\
&\times B_1^e (  t_1, z_1,\mu,\nu)  B_2^f (t_2, z_2, \mu,\nu) J_3^g (p_3^2, \mu,\nu)
J_4^h (p_4^2, \mu,\nu).
\end{align}
Combining all these ingredients,  the 2-jettiness for the final $\mu^- \mu^+$ is obtained by adding 
the contribution from the initial  particles $WW$ and $\ell_e \overline{\ell}_e$, and is factorized as
\begin{align} \label{twojet}
\Bigl(\frac{d\sigma}{d\mathcal{T}_2}\Bigr) &= \frac{1}{2s} \int \frac{dz_1}{z_1} \int\frac{dz_2}{z_2} 
\int dt_1 dt_2 \int \frac{d^4 p_3}{(2\pi)^3} \frac{d^4 p_4}{(2\pi)^3} 
(2\pi)^4 \delta^{(4)} 
\Bigl( \frac{\omega_1 n_1}{2} +\frac{\omega_2 n_2}{2} - \frac{\omega_3 n_3}{2}
-\frac{\omega_4 n_4}{2} \Bigr) \nonumber \\
&\times  \int d\mathcal{T}_s \delta \Bigl( \mathcal{T}_2 - \frac{t_1}{\omega_1} - 
\frac{t_2}{\omega_2} - \frac{p_3^2}{\omega_3} - \frac{p_4^2}{\omega_4} -\mathcal{T}_s\Bigr) 
 \sum_{i=W, e}\sum_{IJ} H_{IJ}^i S_{JI}^{i, efgh} (\mathcal{T}_s , \mu, \nu)
  \nonumber \\
&\times \sum_{efgh} k_i^e k_i^f k_{\mu}^g k_{\mu}^hB_i^e (t_1, z_1, \mu,\nu)  
B_{\bar{i}}^e (t_2, z_2, \mu,\nu) J_{\mu}^g (p_3^2, \mu,\nu)
J_{\bar{\mu}}^h (p_4^2, \mu,\nu),
\end{align}
and the corresponding 
hard and soft functions in each process should be employed.\footnote{Because  the  
contraction of the Dirac structure is different in two channels,  we use $H^e = 16 H$, where $H$ is 
the hard function for $\ell_e \overline{\ell}_e \rightarrow \ell_{\mu} \overline{\ell}_{\mu}$ 
in ref.~\cite{Chay:2021arz}.}
We refer to ref.~\cite{Chay:2021arz} for the factorization in $\mathrm{SCET_{II}}$ in which 
the PDFs are evolved from the scale of order $M$, and matched with the
beam functions at the scale $\sqrt{\mathcal{T}Q}$.

\section{Rapidity divergence\label{rapdi}}
We briefly explain the rapidity divergence and the method of extracting it by prescribing
the rapidity regulators. The rapidity divergence arises 
because the collinear and the soft modes with the same invariant mass reside in disparate phase spaces.
These modes lie on the same hyperbola in the $p^+$-$p^-$ plane,
and they are distinguished only by their rapidities. When the soft particles reach the collinear region, 
the momentum approaches the region $p^+ p^- \rightarrow \infty$, with $p_{\perp}^2$ fixed. 
In this limit,  an additional divergence is induced, which has different source from the UV and the IR
divergences, hence the rapidity divergence cannot be handled with the dimensional regularization. 
The converse is also true. When the collinear particles reach the soft region, the zero-bin subtraction 
should be performed to avoid double counting. And the 
zero-bin contribution induces the rapidity divergence.  However,  no rapidity divergence shows up 
in the full theory because there is no separation of the phase space. It means that there may be rapidity 
divergence in the collinear and the soft sectors in SCET, but it is cancelled in the total contribution.

In spite of the cancellation, we need to regulate the rapidity divergence in each sector. 
There have been many methods
to regulate the rapidity divergence~\cite{collins_2011, Idilbi:2007ff,Idilbi:2007yi, Becher:2011dz,
Chiu:2011qc,Chiu:2012ir, Li:2016axz, Ebert:2018gsn}. Recently,  
one of the authors has constructed  the soft and collinear rapidity regulators in ref.~\cite{Chay:2020jzn}, 
which correctly yields the
angular dependence when different lightcone directions are involved. We will adopt that prescription
to compute the rapidity divergence here. Let us describe
how to prescribe the rapidity regulators in the collinear and the soft sectors.

In the $n$-collinear sector, the basic idea is to attach a rapidity regulator of the form
$(\nu/\overline{n}\cdot k)^{\eta}$ in the $n$-collinear Wilson line, where the rapidity divergence arises
when $\overline{n}\cdot k \rightarrow \infty$ with $\mathbf{k}_{\perp}^2$ fixed.
This prescription is employed in ref.~\cite{Chiu:2012ir}, in which the collinear rapidity divergence 
appears as poles of $\eta$, and $\nu$ is the rapidity scale. The rapidity divergence is manipulated
by modifying the region $\overline{n}\cdot k\rightarrow \infty$.
However, it also modifies the region $\overline{n}\cdot k \rightarrow 0$, which is   
unwanted in the rapidity evolution. The undesirable divergence in this region is cancelled when 
the zero-bin subtraction is performed. 

The construction of the soft rapidity regulator is different from that in ref.~\cite{Chiu:2012ir}, 
and our choice yields the correct angular dependence of the lightlike directions. In terms 
of Feynman diagrams, the $n$-collinear Wilson line is obtained by summing over all the emissions 
of $n$-collinear gluons from the sources other than the $n$-collinear particle. For example, if we consider 
the back-to-back collinear current
$\overline{\xi}_{\bar{n}} W_{\bar{n}} Y^{\dagger}_{\bar{n}} \gamma^{\mu}  Y_n W^{\dagger}_n \xi_n$, 
the $n$-collinear Wilson line is obtained by considering
the emission of the $n$-collinear gluons from the $\overline{n}$-collinear source, and  
it is exponentiated to produce the $n$-collinear Wilson line. 

On the contrary, the soft Wilson line $Y_n$ is obtained by considering the emission of soft gluons from 
the $n$-collinear source. Note that the sources of the emitted gluons to construct the Wilson lines are 
different in the collinear case ($W_n$) and in the soft case ($Y_n$). If we require the consistency to 
treat the rapidity divergence in the collinear and the soft sectors, we should track the same source for 
the rapidity divergence.   Therefore, the 
soft rapidity regulator in the soft Wilson line, corresponding to the $n$-collinear gluons in $W_n$, should be 
attached to the $\overline{n}$-collinear source, that is, $Y_{\bar{n}}$. 
By generalizing the procedure,
let us consider the collinear current $\overline{\xi}_{n_1} W_{n_1}Y_{n_1}^{\dagger}\Gamma
Y_{n_2}W_{n_2}^{\dagger} \xi_{n_2}$, which is invariant under collinear and soft gauge transformations 
respectively. The modified collinear Wilson line $W_{n_1}$ and the soft Wilson line $Y_{n_2}$,
with the rapidity regulators are given as
\begin{align} \label{rareg} 
W_{n_1} &= \sum_{\mathrm{perm.}} \exp \Bigl[
-\frac{g}{\overline{n}_1\cdot \mathcal{P}} \Bigl(\frac{\nu}{|\overline{n}_1\cdot
\mathcal{P}|}\Bigr)^{\eta} \overline{n}_1 \cdot A_{n_1}\Bigr], \nonumber \\ 
Y_{n_2} &=\sum_{\mathrm{perm.}} \exp\Bigl[- \frac{g}{n_2\cdot \mathcal{P}}
\Bigl(\frac{\nu}{|n_2\cdot \mc{P}|} \frac{n_1\cdot n_2}{2}\Bigr)^{\eta} n_2\cdot A_{us}
\Bigr], 
\end{align} 
where $\mathcal{P}$ is the operator extracting the momentum. 
The remaining Wilson lines $W_{n_2}$ and $Y_{n_1}$ can be obtained by switching $n_1$ and $n_2$. 
The point in selecting the rapidity regulator is to trace the same emitted gauge bosons both in the collinear
and the soft sectors, which are eikonalized to produce the Wilson lines. Note that the
rapidity divergences from $W_{n_1}$ and $Y_{n_2}$ have the same origin because the collinear and soft
gauge bosons are emitted from the $n_2$-collinear quark for both of the Wilson lines.

When the rapidity divergence arises in the soft sector, the soft momentum $k$, in the limit 
$\overline{n}_1\cdot k\rightarrow \infty$ approaches $k^{\mu} \approx
(\overline{n}_1 \cdot k) n_1^{\mu}/2$ and the soft rapidity regulator becomes
\begin{equation} 
\Bigl(\frac{\nu}{n_2\cdot k} \frac{n_1\cdot n_2}{2}\Bigr)^{\eta}
\xrightarrow[\bar{n}_1\cdot k \to \infty]{} \Bigl(\frac{\nu}{\overline{n}_1\cdot k}\Bigr)^{\eta},
\end{equation} 
which has the same form as the collinear rapidity regulator for
$W_{n_1}$.  Another pair possessing the same source of rapidity divergence is
$W_{n_2}^{\dagger}$ and $Y_{n_1}^{\dagger}$.

\section{Beam functions, PDF and semi-inclusive jet functions\label{bfpdf}}
 
The bare operator for the gauge-boson beam function is given as
\begin{equation}
\mathcal{O}_W^{a, \mathrm{bare}} (t, \omega) = -\omega \theta (\omega) 
B_{n\perp \mu}^c (0) \delta (t + \omega n \cdot \mathcal{P})
[ \delta (\omega -\overline{n}\cdot \mathcal{P}) B_{n\perp}^{d\mu}(0)] (G^a)^{cd}.
\end{equation}
This operator is extended from QCD~\cite{Berger:2010xi} to 
include the nonsinglets. The beam function for the gauge boson is given by 
the matrix element of the renormalized operator
$\mathcal{O}_W^a (t,\omega, \mu, \nu)$,
\begin{align} \label{wbeam2}
B_W^a (t, x=\omega/P^-,\mu,\nu) &= \langle W (P^-)|\mathcal{O}_W^a (t,\omega, \mu, \nu) 
|W (P^-)\rangle \nonumber \\
&=-\omega\langle W| B_{n\perp \mu}^c \delta (t + \omega n \cdot \mathcal{P})
[ \delta (\omega -\overline{n}\cdot \mathcal{P}) B_{n\perp}^{d\mu}] (G^a)^{cd} |W\rangle.
\end{align}
 Note that there is no nonsinglet beam function for the SU(2) 
gauge group  in the weak interaction because $d^{abc}=0$. However, we will proceed for the general
 SU($N$)  here.
The definition of the singlet beam function for the gauge boson is the same as the gluon beam function in 
QCD~\cite{Berger:2010xi}. However, the authors in ref.~\cite{Berger:2010xi} have computed the gluon 
beam function with massless gluons, with no rapidity regulator.
We compute the radiative corrections of the beam function at NLO with the nonzero gauge boson mass $M$,
and the rapidity regulators.
 
The detailed computation of the beam functions at NLO is relegated to appendix~\ref{nlobf}, and 
we present only the results here.  The singlet and the nonsinglet functions in eq.~\eqref{wbeam2} are denoted 
as $B_{Ws}(G^0)^{ab}$ and $B_{Wn}(G^a)^{bc}$ respectively by extracting the group theory factors.
The singlet beam function for the gauge boson at NLO is given as
\begin{align} \label{bsing}
&B_{Ws}^{(1)} (t, x, \mu) 
  =\frac{\alpha}{2\pi} \Bigl\{ \delta (1-x) \Bigl[ C_A \Bigl( \frac{2}{\eps^2} \delta (t) -
\frac{2}{\eps} \frac{1}{\mu^2} \mathcal{L}_0 \bigl( \frac{t}{\mu^2}\bigr) \Bigr) 
+\frac{\beta_0}{2} \frac{1}{\eps}\Bigr]  \nonumber  \\
&+\delta (t) \Bigl( \frac{\beta_0}{2} \delta (1-x) +C_A P_{WW} (x)\Bigr) \ln \frac{\mu^2}{M^2}
+C_A \delta (t)\Bigl( P_{WW} (x) \ln \frac{1-x}{x} -\frac{\pi^2}{6} \delta (1-x)\Bigr)
 \nonumber \\
&+ C_A \Bigl[ \delta (1-x) \frac{2}{\mu^2} \mathcal{L}_1 \bigl( \frac{t}{\mu^2}\bigr) + P_{WW} (x)
\frac{1}{\mu^2} \mathcal{L}_0 \bigl( \frac{t}{\mu^2}\bigr)\Bigr] \nonumber \\
&+C_A\delta (t) \Bigl[ \delta (1-x) \Bigl( \frac{31}{18} -\frac{\pi^2}{9} -\frac{\pi}{2\sqrt{3}}\Bigr)
-\Bigl(\frac{2(1-x)}{x} + 2x(1-x) +\frac{3}{2}\frac{x(1-x)}{1-x+x^2}\Bigr) \nonumber \\
&-P_{WW} (x) \ln (1-x+x^2)\Bigr] + \delta (1-x) \delta (t) \Bigl[-\frac{2}{9} n_f T_F + \frac{1}{2} n_s T_F
\Bigl(-\frac{17}{9} +\frac{\pi}{\sqrt{3}}\Bigr)\Bigr]\Bigr\}, 
\end{align}
where  $M$ is the mass of the gauge boson.  Compared to  QCD, 
the first three lines in eq.~\eqref{bsing} are the same as those from QCD with the appropriate color factors
after replacing $\ln \mu^2/M^2$ by the IR pole $1/\eir$. The last two lines are the additional contributions due to the nonzero $M$. Here $P_{WW} (x)$ is the splitting function for 
$W\rightarrow WW$, which is the same as the splitting function 
$P_{gg}(x)$ for $g\rightarrow gg$, and is given by
\begin{align}
P_{WW} (x) &= 2\mathcal{L}_0 (1-x) x + 2\theta(1-x) \Bigl[\frac{1-x}{x} + x(1-x)\Bigr] \nonumber \\
&= 2\theta (1-x) \Bigl[ \frac{x}{(1-x)_+} + \frac{1-x}{x} + x(1-x)\Bigr],
\end{align}
and the relation $2(1-x+x^2)^2/x = (1-x)P_{WW}(x)$ is used.
The first term $\beta_0$ in the beta function is given by
\begin{equation}
\beta_0 = \frac{11}{3} C_A - \frac{4}{3} \Bigl(\frac{1}{2}\Bigr)  n_f T_f -\frac{1}{3} n_s T_f,
\end{equation}
where $n_s$ is the number of complex scalar multiplets in the theory\footnote{In the standard model with the 
SU(2) weak interaction, there is one scalar multiplet, but in SU($N$) gauge theory we need more multiplets
to attain equal masses $M$ of the gauge bosons. We will not dwell on the detailed model building to achieve this
here.},  and the factor 1/2 in front of $n_f$ evokes the fact that
only the left-handed fields contribute here.
 
The nonsinglet beam function for the gauge boson at NLO is given as
\begin{align} \label{bnon}
&B_{Wn}^{(1)} (t,x,\mu,\nu) 
 = B_{Ws}^{(1)} (t, x, \mu) \nonumber \\
&+\frac{\alpha C_A}{4\pi}\Bigl\{ 2\delta (1-x) \Bigl[ \delta (t) \Bigl( \frac{1}{\eps} 
+ \ln \frac{\mu^2}{M^2}\Bigr) \Bigl( \frac{1}{\eta}
+\ln \frac{\nu}{p^-}\Bigr) -\frac{1}{\eps^2} \delta (t)
+ \frac{1}{\eps} \frac{1}{\mu^2} \mathcal{L}_0 \bigl( \frac{t}{\mu^2}\bigr)  -\frac{1}{\mu^2}\mathcal{L}_1 
\bigl(\frac{t}{\mu^2}\bigr)\Bigr] \nonumber \\
&-P_{WW} (x)\Bigl( \delta (t) \ln \frac{\mu^2}{M^2} +\frac{1}{\mu^2} \mathcal{L}_0 
\bigl( \frac{t}{\mu^2}\bigr)  \Bigr) +\delta (t) \Bigl( P_{WW}(x)  \ln \frac{x(1-x+x^2)}{1-x}+\frac{\pi^2}{6}
\delta(1-x)   \nonumber \\
&+   2\frac{1-x}{x} + 2x(1-x) +\frac{3}{2} \frac{x(1-x)}{1-x+x^2}\Bigr) \Bigr\}.
\end{align}
The result  for the nonsinglet beam function in eq.~\eqref{bnon} is new. Note that the nonsinglet 
beam function contains the rapidity divergence.

In our approach with \scone, the beam functions are convoluted with the jet, the soft and the hard
functions. It corresponds to the case, in which the hierarchy of the scales satisfies $\mathcal{T}^2 \sim M^2 \ll
p_c^2 \sim Q\mathcal{T} \ll Q^2$. However, there can be another situation in which $p_c^2 \ll Q\mathcal{T}$. 
  In this case, the previous collinear momentum in \scone,
scaling as $(Q,\sqrt{Q\mathcal{T}}, \mathcal{T})$ is labeled as the hard-collinear momentum, and the corresponding
degrees of freedom should be integrated out to yield \sctwo. The collinear momentum in \sctwo\ scales as $p_c^{\mu}
\sim (Q, \mathcal{T}, \mathcal{T}^2/Q)$. In this procedure the beam function is matched onto the PDF,
with the appropriate matching coefficients.

In \sctwo, the PDF near the scale $\mathcal{T} \sim M$ is evolved to the hard-collinear scale, and is matched
to the beam functions. And the beam functions are evolved to the factorization scale. The two cases in \scone\ and
\sctwo\ are considered in detail in ref~\cite{Chay:2021arz}. We can deal with both cases, but for simplicity
we consider the case of \scone\ only. Therefore  the matching coefficients between the beam 
functions and the PDF below are derived to be used in \sctwo\ for completeness, but they will not be considered here.

The bare gauge-boson PDF operator is defined as
\begin{equation}
\mathcal{Q}_W^{a, \mathrm{bare}} (\omega) = -\omega \theta(\omega) 
B_{n\perp \mu}^c (0)  [ \delta (\omega -\overline{n}\cdot \mathcal{P}) B_{n\perp}^{d\mu}(0)] 
(G^a)^{cd},
\end{equation}
and the PDF for the gauge boson is given by the matrix element as
\begin{equation} \label{wpdf2}
f_W^a (x=\omega/P^-,\mu,\nu) = \langle W (P^-)|\mathcal{Q}_W^a (\omega, \mu, \nu) |W (P^-)\rangle.
\end{equation}
By performing the operator-product expansion of $\mathcal{O}_W^a (t,\omega, \mu, \nu)$, it can 
be expressed in
terms of the operators $\mathcal{Q}_W^b (\omega)$ with the matching coefficients 
$\mathcal{I}_{ij}^{ab}$ as
\begin{equation}
\mathcal{O}_W^a (t,\omega, \mu,\nu) = \sum_{j,b} \int\frac{d\omega'}{\omega'} \mathcal{I}_{Wj}^{ab}  
\Bigl( t,\frac{\omega}{\omega'}, \mu\Bigr) \mathcal{Q}_j^b (\omega',\mu, \nu), 
\end{equation}
to leading order in SCET. By taking the matrix elements, we obtain the relation between the beam function and 
the PDF as 
\begin{equation} \label{beammat}
B_W^a (t, x,  \mu, \nu) =\sum_{j,b} \int_x^1 \frac{dx'}{x'} \mathcal{I}_{Wj}^{ab} 
\Bigl( t, \frac{x}{x'}, \mu\Bigr) f_j^b (x', \mu, \nu).
\end{equation} 
The matching coefficients $\mathcal{I}_{ij}^{ab}$ describe the collinear initial-state radiation
and can be computed perturbatively.  Note that they are independent of the rapidity scale $\nu$. 
Here $i$, $j$ are the indices for particle species, and $a$, $b$ 
are the weak indices.   

The singlet and the nonsinglet PDFs for the gauge bosons are denoted as $f_{Ws}(G^0)^{ab}$ and
$f_{Wn} (G^a)^{bc}$ respectively. At NLO, the singlet PDF is given as 
\begin{align} \label{fsing}
f_{Ws}^{(1)}  
&= \frac{\alpha}{2\pi}   \Bigl\{  \Bigl(C_A P_{WW} (x) +\frac{\beta_0}{2} \delta (1-x)\Bigr) \Bigl( \frac{1}{\eps}
+ \ln \frac{\mu^2}{M^2}\Bigr)  \nonumber \\
& -C_A \Bigl(P_{WW} (x) \ln (1-x+x^2) +2\frac{1-x}{x} +2x(1-x) +\frac{3}{2} \frac{x(1-x)}{1-x+x^2} 
\Bigr)   \nonumber \\
& +\delta (1-x) \Bigl[ C_A \Bigl( \frac{31}{18} -\frac{\pi}{2\sqrt{3}} -\frac{\pi^2}{9}\Bigr) 
- \frac{2}{9} n_f   T_F   
 + \frac{1}{2}n_s T_F \Bigl( -\frac{17}{9} +\frac{\pi}{\sqrt{3}}  \Bigr) \Bigr\}. 
\end{align}
 The matching coefficient $\mathcal{I}_{WW}^{s(1)}$ is obtained by comparing
 eqs.~\eqref{bsing} and \eqref{fsing}  as
\begin{align} \label{iww}
\mathcal{I}_{WW}^{s(1)}  &= \frac{\alpha C_A}{2\pi} \Bigl\{ 
\frac{2}{\mu^2} \mathcal{L}_1 \Bigl( \frac{t}{\mu^2}\Bigr) \delta (1-x) +\frac{1}{\mu^2}
 \mathcal{L}_0 \Bigl( \frac{t}{\mu^2}\Bigr) P_{WW} (x) \nonumber \\
&    +\delta (t) \Bigl[  P_{WW}(x)  \ln \frac{1-x}{x}   
 -\frac{\pi^2}{6}  \delta (1-x)\Bigr]  \Bigr\}.
 \end{align}
It is the same as the matching coefficient for the gluon case in QCD except the group theory factor. 
(See eq.~(2.14) in ref.~\cite{Berger:2010xi}.) 
 
The nonsinglet PDF for the gauge boson at NLO  is given as
\begin{align}
f_W^{n(1)}   &= f_W^{s(1)}   +\frac{\alpha}{2\pi}\frac{C_A}{2}   \Bigl[ \delta(1-x)  
\Bigl(\frac{2}{\eta} +2\ln \frac{\nu}{p^-}\Bigr) \Bigl( \frac{1}{\eps} +\ln \frac{\mu^2}{M^2}\Bigr)  
\nonumber \\
&-  \Bigl(\frac{1}{\eps} +\ln \frac{\mu^2}{M^2}\Bigr)  P_{WW} (x) +2\frac{1-x}{x} +2x (1-x)
+\frac{3}{2} \frac{x(1-x)}{1-x+x^2} \nonumber \\
&+P_{WW} (x) \ln (1-x+x^2)\Bigr].
\end{align}
The matching coefficient for the nonsinglet $\mathcal{I}_{WW}^{n(1)}$ is proportional to the singlet
matching coefficient $\mathcal{I}_{WW}^{s(1)}$, and its relation is given by
\begin{equation}
\mathcal{I}_{WW}^{n(1)} =\frac{1}{2} \mathcal{I}_{WW}^{s(1)}.
\end{equation}
 It is interesting to note that the
matching coefficients for the nonsingets are proportional to those for the singlets both for the gauge bosons
and for the leptons though the proportionality constant is different. 
 $I_{\ell\ell}^{n(1)} =-I_{\ell\ell}^{s(1)}/(N^2-1)$.  (See ref.~\cite{Chay:2021arz}.)
 
For completeness, we present  the lepton beam functions from ref.~\cite{Chay:2021arz}.
We express the singlet and nonsinglet beam functions $B_{\ell}^0$ and $B_{\ell}^a$ in terms of 
$B_s$ and $B_n$ by 
extracting and separating the group theory factors as 
\begin{equation} \label{lbeamel}
B^0_{\ell} (t,x,  M, \mu) = B_s (t, x, M, \mu) \mathrm{Tr} (T^0 P_{\ell}), \ 
B^a_{\ell} (t, x, M, \mu) = B_n (t, x, M, \mu) \mathrm{Tr} (T^a P_{\ell}).
\end{equation}
The bare beam functions $B_s$ and $B_n$ are given at NLO as
\begin{align}
B_s^{(1)} (t, x, M, \mu)  
&=\frac{\alpha C_F}{2\pi} \Bigl\{  \delta (t) \delta (1-x) \Bigl( \frac{2}{\eps^2} +\frac{3}{2\eps}  
+\frac{9}{4} -\frac{\pi^2}{2}\Bigr)   \\
&+\delta (t) \Bigl[ P_{\ell \ell}(x) \ln \frac{\mu^2}{x^2 M^2}  + (1+x^2) \mathcal{L}_1 (1-x) -(1-x) \theta (x)
\theta (1-x) \Bigr]\nonumber \\
& + \delta (1-x) \Bigl[ -\frac{2}{\eps}  \frac{1}{\mu^2} \mathcal{L}_0
\Bigl(\frac{t}{\mu^2}\Bigr) +\frac{2}{\mu^2} \mathcal{L}_1
\Bigl(\frac{t}{\mu^2}\Bigr) \Bigr]  + (1+x^2) \mathcal{L}_0 (1-x) 
\frac{1}{\mu^2} \mathcal{L}_0 \Bigl( \frac{t}{\mu^2}\Bigr)\Bigr\}, \nonumber \\
B_n^{(1)} (t, x, M, \mu)  &= B_{s} (t, x, M, \mu) \nonumber \\
& -\frac{\alpha C_A}{4\pi} \Bigl\{  -2 \delta (t) \delta (1-x) \Bigl[ \Bigl(\frac{1}{\eta} 
+\ln \frac{\nu}{\omega}\Bigr)\Bigl(
\frac{1}{\eps} +\ln \frac{\mu^2}{M^2}\Bigr) -\frac{1}{\eps^2} +\frac{\pi^2}{12} \Bigr] \nonumber \\ 
&+ \delta (t) \Bigl[ (1+x^2) \mathcal{L}_0 (1-x) \ln \frac{\mu^2}{x^2 M^2} + (1+x^2) \mathcal{L}_1 (1-x)
-(1-x) \theta (x) \theta (1-x)\Bigr] \nonumber \\
&+\delta (1-x) \Bigl[-\frac{2}{\eps} \frac{1}{\mu^2} \mathcal{L}_0
\Bigl(\frac{t}{\mu^2}\Bigr)  +\frac{2}{\mu^2} \mathcal{L}_1 \Bigl(\frac{t}{\mu^2}\Bigr) \Bigr]
+ (1+x^2) \mathcal{L}_0 (1-x) \frac{1}{\mu^2} \mathcal{L}_0
\Bigl(\frac{t}{\mu^2}\Bigr) \Bigr\}. \nonumber
\end{align} 
Here the splitting function $P_{\ell \ell }(x)$ for $\ell \rightarrow \ell W$ is the same as the quark
splitting function $P_{qq} (x)$, and is given by 
\begin{equation} 
P_{\ell\ell} (x) = P_{qq} (x) = \mathcal{L}_0 (1-x) (1+x^2) +\frac{3}{2}\delta (1-x) = \Bigl[ \theta (1-x)
\frac{1+x^2}{1-x}\Bigr]_+. 
\end{equation} 


The bare singlet and nonsinglet semi-inclusive jet functions $J_l^0$ and $J_l^a$, 
with the lepton $l$ in the final state, are given as~\cite{Chay:2021arz}
\begin{equation} \label{jetex}
J_l^0 (p^2 , M, \mu) =  J_s (p^2, M, \mu) \mathrm{Tr} (P_l T^0), \ J_l^a (p^2 , M, \mu) = 
 J_n (p^2, M, \mu) \mathrm{Tr} (P_l T^a),
\end{equation}
where $P_l$ is the projection operator to the lepton $l$.
The bare singlet and the nonsinglet jet functions at NLO are given as 
\begin{align} 
J_s^{(1)} (p^2, M, \mu)  
&=\frac{\alpha C_F}{2\pi} \Bigl[ \delta (p^2) \Bigl( \frac{2}{\eps^2} +\frac{3}{2\eps}
+\frac{7}{2} -\frac{\pi^2}{2}\Bigr) -\Bigl( \frac{2}{\eps}
+\frac{3}{2}\Bigr)\frac{1}{\mu^2} \mathcal{L}_0 \Bigl(\frac{p^2}{\mu^2}\Bigr)
+\frac{2}{\mu^2} \mathcal{L}_1 \Bigl(\frac{p^2}{\mu^2}\Bigr) \Bigr], \nonumber \\
J_n^{(1)} (p^2, M, \mu)  &= J_s (p^2, M, \mu) + \frac{\alpha
C_A}{2\pi}\Bigl\{ \delta (p^2) \Bigl[\frac{1}{\eta}\Bigl( \frac{1}{\eps} +\ln
\frac{\mu^2}{M^2}\Bigr) -\frac{1}{\eps^2} +\frac{1}{\eps} \ln \frac{\nu}{\omega} + \ln
\frac{\mu^2}{M^2} \ln \frac{\nu}{\omega} \nonumber \\ 
&+\frac{3}{4} \ln \frac{\mu^2}{M^2}
-\frac{5}{8} +\frac{\pi^2}{12}\Bigr] + \Bigl( \frac{1}{\eps} +\frac{3}{4}\Bigr)
\frac{1}{\mu^2} \mathcal{L}_0 \Bigl(\frac{p^2}{\mu^2}\Bigr) -\frac{1}{\mu^2} \mathcal{L}_1
\Bigl(\frac{p^2}{\mu^2}\Bigr) \Bigr\}.   
\end{align} 
Like the beam functions, the nonsinglet jet function develops the rapidity divergence. 

\section{Hard function\label{hardf}}
 
The hard functions can be read off from those in QCD~\cite{Kelley:2010fn}. However, the bases in 
ref.~\cite{Kelley:2010fn} are $T'_1=t^a t^b$, $T'_2=t^b t^a$ and $T'_3=\delta^{ab}$, while our bases 
consist of $T_1=\delta^{ab}$, $T_2=if^{abc}t^c$ and $T_3= d^{abc}t^c$. The change of basis can be obtained 
by noting the relation
\begin{equation}
t^a t^b = \frac{1}{2}\Bigl[ \frac{1}{N} \delta^{ab} +(if^{abc} +d^{abc}) t^c \Bigr].
\end{equation}
Therefore the relation between $T'_i$ and $T_i$ is given by 
\begin{equation}
(T'_1 \ T'_2 \  T'_3) =   (T_1\  T_2 \ T_3) \mathbf{A},
\end{equation}
where the transformation matrix $\mathbf{A}$ is given by
\begin{equation}
\mathbf{A}= \begin{pmatrix}
1/2N & 1/2N &1 \\
1/2 & \displaystyle -1/2 & 0\\
1/2& 1/2 & 0\end{pmatrix}.
\end{equation}
Then the sum of the operators $D'_I O'_I$ of the effective Lagrangian in eq.~\eqref{effl} has the relation
$O'_I D'_I = O_I A_{IJ} D'_J = O_I D_I$, from which we obtain the relation $D_I = A_{IJ} D'_J$. Since
the hard coefficients $H_{IJ}$ are proportional to $D_I D^*_J$, $\mathbf{H} = \mathbf{A}
\mathbf{H}' \mathbf{A}^{\dagger}$. 

In addition to the change of the basis, the appropriately adjusted color factors for SU(2), and the fact 
that only the left-handed leptons contribute to our hard functions should be implemented. With this 
in mind, the hard function can be written as
\begin{equation}
H_{IJ}(s,t,u) = 4g^4 H_{IJ}^{(0)} + 8g^4 \frac{\alpha}{4\pi} H_{IJ}^{(1)} +\cdots,
\end{equation}
where $H_{IJ}^{(0)}$ is the hard function at leading order (LO), and $H_{IJ}^{(1)}$  at NLO. 
The Mandelstam variables $s$, $t$, $u$  are given by $s =(p_1 + p_2)^2 = (p_3+p_4)^2$, 
$t=  (p_1 -p_3)^2 = (p_2 - p_4)^2$,  $u =  (p_1 -p_4)^2 = (p_2 - p_3)^2$, where $p_i$ are the partonic 
momenta. The variable $u$ is given by
$u = -\omega_1 \omega_4 n_1 \cdot n_4/2 = -\omega_2 \omega_3 n_2 \cdot n_3/2$, and $s$, $t$ can be
expressed accordingly.

The LO hard coeffcients $H_{IJ,\mathrm{QCD}}^{\prime(0)}$ from ref.~\cite{Kelley:2010fn} are given as
\begin{equation}
\mathbf{H}^{\prime (0)} = \frac{1}{s^2}\begin{pmatrix}
\displaystyle \frac{u}{t} (t^2 +u^2) & \displaystyle  t^2 +u^2  &0\\
\displaystyle  t^2 +u^2  &  \displaystyle \frac{t}{u}(t^2 +u^2)& 0\\
0&0&0
\end{pmatrix},
\end{equation}
and in our basis it becomes
\begin{equation}
\mathbf{H}_W^{(0)} = \frac{1}{4s^2tu}\begin{pmatrix}
  (t+u)^2 (t^2 +u^2)/N^2  &   (u^4 -t^4)/N  & (t+u)^2 (t^2 +u^2)/N \\
(u^4 -t^4)/N &  (t-u)^2 (t^2+u^2)& u^4 -t^4\\
(t+u)^2 (t^2 +u^2)/N&u^4 -t^4&(t+u)^2 (t^2 +u^2)
\end{pmatrix}. 
\end{equation}

The  NLO hard functions $H^{\prime(1)}_{IJ}$ are given as
\begin{align}
H_{11}^{\prime (1)} &= \frac{u(t^2 +u^2)}{ts^2} \Bigl[ -(C_A+C_F) (L(s))^2 +V_1 (s,t,u) L(s)\Bigr]
+\frac{tu}{s^2} W_1 (s,t,u) + \frac{u^3}{ts^2} W_2 (s,t,u), \nonumber \\
H_{22}^{\prime (1)} &= \frac{t(t^2 +u^2)}{us^2} \Bigl[ -(C_A+C_F) (L(s))^2 +V_1 (s,u,t) L(s)\Bigr]
+\frac{tu}{s^2} W_1 (s,u,t) + \frac{t^3}{us^2} W_2 (s,u,t), \nonumber \\
H_{12}^{\prime (1)} &= \frac{t^2 +u^2}{2s^2} \Bigl[ -2 (C_A+C_F) (L(s))^2 + V_1 (s,t,u) L(s) 
+V_1 (s, u,t) L(s) \Bigr]   \\
&+ \frac{u^2}{2s^2} \Bigl( W_1 (s,u,t) +W_2 (s,t,u)\Bigr) +\frac{t^2}{2s^2} 
\Bigl( W_1 (s,t,u) + W_2 (s,u,t)\Bigr), \nonumber \\
H_{13}^{\prime (1)} &= \frac{t}{2s} V_2 (s,t,u) L(s) +\frac{u^2}{2st} L(s) +\frac{t}{2s} W_4 (s,t,u) 
+\frac{u^2}{2st} W_4 (s,u,t), \nonumber \\
H_{23}^{\prime (1)} &= \frac{t^2}{2su} V_2 (s,t,u) L(s) +\frac{u}{2s} V_2 (s,u,t) L(s) +\frac{t^2}{2su}
W_4 (s,t,u) +\frac{u}{2s} W_4 (s,u,t), \ \ H_{33}^{\prime (1)} =0, \nonumber
\end{align}
and $H'_{JI} = H'^{*}_{IJ}$.  The functions $V_i$ and $W_i$ are given by
\begin{align}
W_1 (s,t,u) &= (C_A -C_F) \frac{s}{u} \Bigl[ \Bigl( L(s) -L(t)\Bigr)^2 + \pi^2\Bigr] +C_A -8C_F
+ (7C_A +C_F)\frac{\pi^2}{6},   \\
W_2 (s,t,u)&= \Bigl( -C_F \frac{s^3}{u^3} -C_A \frac{t^3 +u^3 -s^3}{2u^3} \Bigr) \Bigl[ \Bigl(
L(s) -L(t)\Bigr)^2 +\pi^2\Bigr] \nonumber \\
&+\Bigl( 2C_A \frac{ts}{u^2} +C_F \frac{s(2s-u)}{u^2} \Bigr) \Bigl( L(t) -L(s)\Bigr) +C_F \frac{t-7u}{u}
-C_A \frac{t}{u} + (7C_A +C_F) \frac{\pi^2}{6}, \nonumber \\
W_3 (s,t,u) &= 2C_F -2C_A -\frac{2t}{3s} (C_A -n_f), \nonumber \\
W_4 (s,t,u) &= -\frac{3u}{4t} \Bigl(L(s) - L(u)\Bigr)^2 -\Bigl( L(s) -L(t)\Bigr) \Bigl( L(s) - L(u)
\Bigr) +\frac{3\pi^2}{2} \frac{u^2}{ts}, \nonumber \\
V_1 (s,t,u) &= 3C_F -2C_A \Bigl( L(t) -L(s)\Bigr) +\beta_0 \ \ V_2 (s,t,u) = \Bigl( L(s) - L(u)\Bigr)
+\frac{t}{s} \Bigl( L(t) -L(u)\Bigr). \nonumber
\end{align}
The function $L(x)$ as a function of the Mandelstam variables is given by 
\begin{equation} 
L(t) = \ln \frac{-t}{\mu^2}, \ L(u) = \ln \frac{-u}{\mu^2}, \ L(s) = \ln \frac{s}{\mu^2} -i\pi. 
\end{equation} 
In order to obtain $H_{IJ}^{(1)}$ in our basis, we perform the transformation $\mathbf{H} = \mathbf{A}
\mathbf{H}' \mathbf{A}^{\dagger}$. 
 
\section{Soft function\label{softf}}
The soft function from $WW\rightarrow \ell_{\mu} \overline{\ell}_{\mu}$ is written in 
terms of  $3\times 3$ matrices in the basis of the operators in eq.~\eqref{opbasis}. The computation for 
the virtual and the real contributions is the same as that in $\ell_e \overline{\ell}_e 
\rightarrow \ell_{\mu} \overline{\ell}_{\mu}$, but
the complexity comes from the group theory factors in the soft Wilson lines. [See eq.~\eqref{softdef}.] 
The number of the color factors to be computed is of the order of 6000 to NLO for 
$WW\rightarrow \ell_{\mu} \overline{\ell}_{\mu}$, while it is of order 1000 for 
$\ell_e \overline{\ell}_e \rightarrow \ell_{\mu} \overline{\ell}_{\mu}$. We have constructed a 
Mathematica package to compute all the group theory factors. 

\begin{figure}[b] 
 \begin{center}
\includegraphics[height=5.8cm]{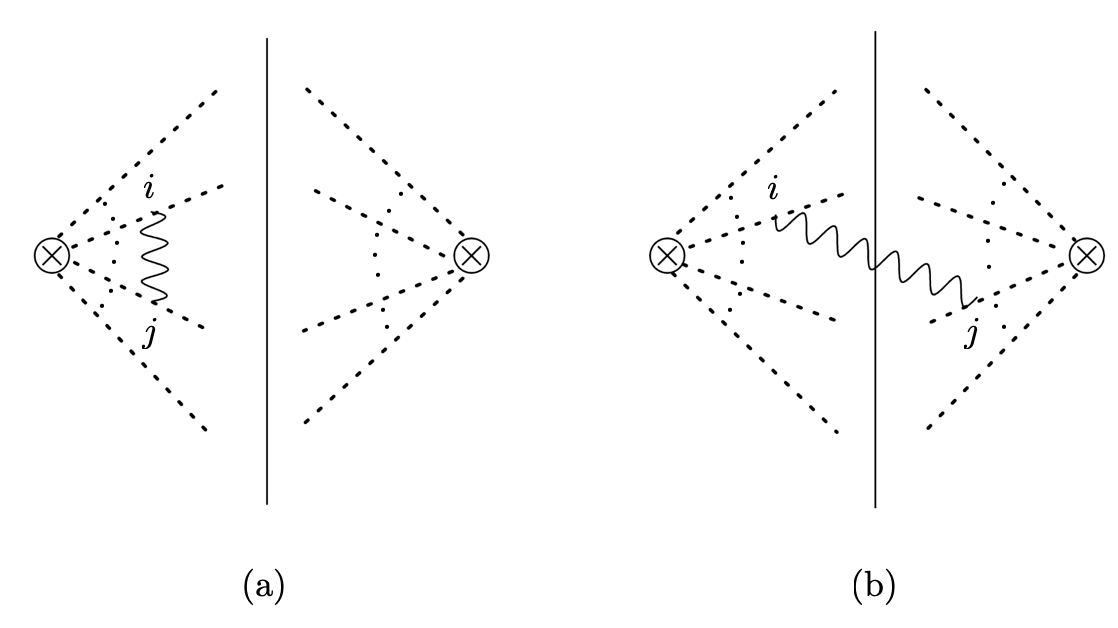} 
\end{center} \vspace{-0.6cm}
\caption{\label{softwilson}\baselineskip 3.0ex  Feynman diagrams for the emission of a soft gauge
boson from the soft Wilson lines $Y$ or $\mathcal{Y}$. The vertical lines are the final-state cut.  (a)
The virtual contribution,   (b)  the real contribution.  
The gauge bosons attached to the same $i$ do not contribute because $n_i^2 = 0$. All the possible
pairs of contractions with different $i$ and $j$ should be summed.} 
\end{figure}

The relevant Feynman diagrams are shown in fig.~\ref{softwilson}. The virtual contributions in 
fig.~\ref{softwilson}(a) are obtained by contracting different soft Wilson lines on the same side of the 
unitarity cut. The contraction between the same Wilson lines vanishes because $n_i^2 =0$.  
The real contribution, shown in fig.~\ref{softwilson}(b) is obtained by contracting the soft Wilson
lines across the unitary cut. 

As in $\ell_e \overline{\ell}_e \rightarrow \ell_{\mu} \overline{\ell}_{\mu}$, 
we decompose the soft function into the hemisphere and the non-hemisphere parts. 
We implement the hemisphere function by the constraint function $F$,
with four independent labels, as~\cite{Chay:2021arz}
\begin{equation} \label{softjetti}
F(k, \{ q_i \} ) = F_{ij, \mathrm{hemi}} (k, \{q_i \}) + F_{ij, ml} (k, \{q_i \}) 
+  F_{ij, lm} (k, \{q_i \}) + (i \leftrightarrow j),
\end{equation}
where the hemisphere measurement function for the full hemisphere $q_j > q_i$ is given by
\begin{equation}
F_{ij, \mathrm{hemi}} (k, \{q_i \}) = \theta(q_j - q_i)  \delta (k-q_i).
\end{equation}
The indices $l$, $m$ refer to the remaining lightcone directions when we contract the Wilson lines in the
$i$ and $j$ directions.
The non-hemisphere functions are given as
\begin{align}
F_{ij, ml} (k, \{q_i \}) &= \theta (q_j - q_i)\theta (q_i - q_m) \theta (q_l - q_m) 
\Bigl( \delta (k-q_m) -\delta (k-q_i)\Bigr),  \nonumber \\
F_{ij, lm} (k, \{q_i \}) &= \theta (q_j - q_i) \theta (q_i -q_l) \theta (q_m - q_l) 
\Bigl( \delta (k-q_l) -\delta (k-q_i)\Bigr),
\end{align}
which are the non-hemisphere measurement function for the regions $m$ and $l$ respectively.
Note that the constraint function $F$ is constructed for the 
gauge boson emitted from the soft Wilson lines  $Y_i$ ($\mathcal{Y}_i$) and $Y_j^{\dagger}$ 
($\mathcal{Y}_j$). 
The hemisphere function for the $i$ and $j$ jet directions contains the collinear
and the soft divergences.  
It is enough to focus on the hemisphere soft function, in which all the divergences reside. We can 
obtain the relevant anomalous dimensions at NLL. 
 
We cite the hemisphere virtual contributions $S_{ij,\mathrm{hemi}}^V$ and the real contributions
$S_{ij,\mathrm{hemi}}^V$ from ref.~\cite{Chay:2021arz} as
\begin{align}
S_{ij,\mathrm{hemi}}^V &=\frac{\alpha}{2\pi} \delta (k)   \Bigl[ \frac{1}{\eps^2} -\frac{2}{\eta}
\Bigl( \frac{1}{\eps} +\ln \frac{\mu^2}{M^2}\Bigr)
-\frac{1}{\eps} \ln \frac{n_{ij} \nu^2}{\mu^2} 
+\frac{1}{2} \ln^2 \frac{\mu^2}{M^2} - \ln \frac{n_{ij} \nu^2}{M^2}\ln \frac{\mu^2}{M^2}
-\frac{\pi^2}{12}\Bigr],  \nonumber \\
S_{ij,\mathrm{hemi}}^{R} &=
-\frac{\alpha}{2\pi} \Bigl\{ \delta (k) \Bigl[ \frac{2}{\eta}\Bigl( \frac{1}{\eps} +\ln
\frac{\mu^2}{M^2}\Bigr) -\frac{2}{\eps^2} +\frac{1}{\eps} \ln \frac{\nu^2}{\mu^2} 
+\ln \frac{\nu^2}{M^2} \ln \frac{\mu^2}{M^2} +\frac{\pi^2}{6}\Bigr]  \\ 
&+ \Bigl(
\frac{1}{\eps} +\ln \frac{\mu^2}{M^2}\Bigr) \frac{2}{\mu} \mathcal{L}_0 \Bigl(
\frac{k}{\mu}\Bigr) \theta (k< \sqrt{n_{ij}} M) + \frac{2}{k} \Bigl( \frac{1}{\eps} + \ln
\frac{n_{ij} \mu^2}{k^2} \Bigr) \theta (k > \sqrt{n_{ij}} M)\Bigr\},  \nonumber
\end{align}
where $n_{ij} = n_i \cdot n_j/2$.
In terms of these contributions, the hemisphere soft function can be written as
\begin{equation}
\mathbf{S}_{\mathrm{hemi}} (a_1, a_2, a_3, a_4) = \sum_{i\neq j} \Bigl[ \mathbf{S}_{ij}^V (a_1, a_2, a_3, a_4) 
S_{ij, \mathrm{hemi}}^V + \mathbf{S}_{ij}^R (a_1, a_2, a_3, a_4) S_{ij, \mathrm{hemi}}^R \Bigr].
\end{equation}
The factors $\mathbf{S}_{ij}^{V,R} (a_1, a_2, a_3, a_4)$ in front of the virtual and real contributions 
represent the corresponding color factors, and they are represented in terms of the $3\times 3$ matrices 
for $WW\rightarrow\ell_{\mu} \overline{\ell}_{\mu}$. The indices $a_i$ denote the color index in 
the presence of the nonsinglets from the originating $i$-th collinear particle  ($i=1,2$  
for the incoming particles, and $i=3, 4$ for the 
outgoing particles  in our convention).  For example, the soft color matrix  
with all the singlets is given 
by $\mathbf{S}(0,0,0,0)$ and the soft color matrix with the 
nonsinglet contributions from 1 and 3 is denoted as $\mathbf{S} (1,0,1,0)$, etc.. All the soft matrices
at tree level are presented in appendix~\ref{softzm}.

\section{Anomalous dimensions and RG evolution\label{anomd}}

The factorized hard, collinear and soft parts contain the logarithms which become small at
their own characteristic scales. The typical hard scale is  $\mu_H\sim Q$, and the collinear scale is 
given by $\mu_C \sim \sqrt{Q\mathcal{T}} \sim \sqrt{QM}$, while the soft scale is 
$\mu_S \sim M$. However, if the common factorization scale  $\mu_F$ is away from these 
characteristic scales, the logarithms become so large that the perturbation theory breaks down. We can resum
large logarithms by evolving the factorized parts from their characteristic scales to the common 
factorization scale $\mu_F$ by solving the RG equations. 
 
\begin{figure}[b] 
\centering
\includegraphics[height=6.cm]{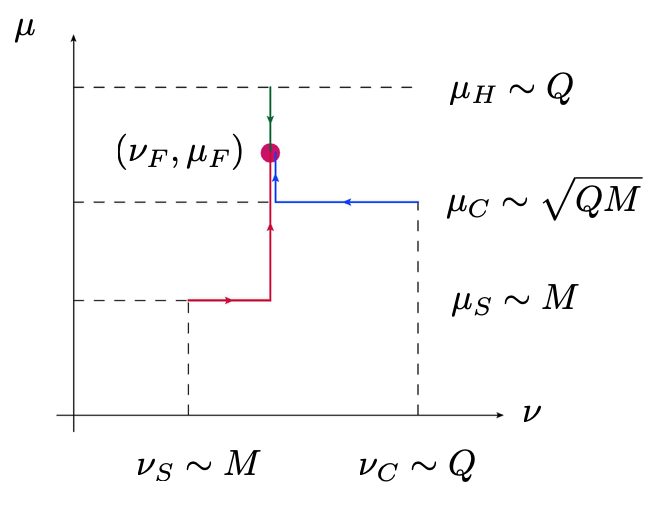} 
\caption{\label{evolution}\baselineskip 3.0ex  The evolutions of the hard, collinear and soft functions start 
from their own characteristic scales $\mu_H$, ($\mu_C$, $\nu_C$), and ($\mu_S$, $\nu_S$) respectively. 
Since the order of the evolution is irrelevant, we first evolve the collinear and soft functions with respect 
to $\nu$, and then $\mu$. The path of the evolution is illustrated by the arrows. } 
\end{figure}

The nonsinglet contributions contain additional logarithms associated with the rapidity divergence. The 
characteristic  collinear and soft rapidity scales are  $\nu_C \sim Q$,  $\nu_S \sim M$ respectively. 
We perform the double evolution with respect to the rapidity scale $\nu$,
as well as the renormalization scale $\mu$. 
Because the order of the evolution is irrelevant~\cite{Chiu:2012ir}, we first evolve with respect 
to $\nu$ from $\nu_C \sim Q$ for the collinear functions 
and from $\nu_S\sim M$ for the soft functions to the common factorization scale $\nu_F$. Then
we evolve all the factorized functions with respect to $\mu$. The path of the evolution is shown in
fig.~\ref{evolution}.  Note that the collinear scale starts at $\mu_c \sim \sqrt{QM}$ in \scone, and
we use the evolution of the beam functions from $\mu_C$ to the factorization scale $\mu_F$. However,
if we employ \sctwo, in which we employ the PDF, the evolution of the PDF should start from the scale 
$\mu_S \sim M$ to $\sqrt{QM}$.  Above the scale $\sqrt{QM}$, the evolution of the beam functions should be 
employed after the matching.

\subsection{Collinear functions}
There are various collinear functions involved in the 2-jettiness. They are the beam functions for the
leptons and the gauge bosons, the corresponding PDFs , and the semi-inclusive muon jet functions. 
Because of the crossing symmetry, the anomalous dimensions
of the beam functions and the corresponding jet functions turn out to be the same. 

It is convenient to express the $N$-jettiness in terms of the Laplace transforms, in which the 
factorization consists of the products instead of the convolution. The actual $N$-jettiness is obtained by
taking the inverse Laplace transform, which will be discussed more in the numerical estimates.
The Laplace transforms for  the beam functions are written as
\begin{align} \label{lapcol2}
\tilde{B}_i \Bigl( \ln \frac{\omega Q_L}{\mu^2} ,z, M, \mu \Bigr) &= \int_0^{\infty} dk \, e^{-s k} 
B_i (\omega k, z, M, \mu), \nonumber \\
 B_i (\omega k, z, M, \mu) &= \frac{1}{2\pi i} \int_{c-i\infty}^{c+i\infty} ds 
\, e^{s k} \tilde{B}_i \Bigl( \ln \frac{1}{e^{\gamma_{\mathrm{E}}} s \mu }, M, \mu \Bigr), 
\end{align}
with $s=1/(e^{\gamma_{\mathrm{E}}} Q_L)$. Here $Q_L$ is the scale introduced in performing the
Laplace transform. The Laplace transforms of the PDFs and the jet functions are defined similarly.
 
Let us denote $\tilde{f}_s$ and $\tilde{f}_n$ as the singlet and the nonsinglet Laplace-transformed 
collinear functions, which can be either the beam functions, or 
the semi-inclusive muon jet functions. The singlet
functions evolve with respect to the renormalization scale $\mu$ only, and the RG equation is written as
\begin{equation} \label{singev}
\frac{d}{d\ln\mu} \tilde{f}_s (\mu) = \tilde{\gamma}_{fs}\tilde{f}_s (\mu),
\end{equation} 
and  the nonsinglet functions $\tilde{f}_n$ satisfy the coupled RG equations, which are 
given as
\begin{equation} \label{nonev}
\frac{d}{d\ln \mu} \tilde{f}_n (\mu,\nu) = \tilde{\gamma}_{fn}^{\mu} \tilde{f}_n (\mu,\nu), \ \ 
\frac{d}{d\ln \nu} \tilde{f}_n(\mu,\nu) = \tilde{\gamma}_{fn}^{\nu} \tilde{f}_n (\mu,\nu).
\end{equation} 
Here $\tilde{\gamma}_{fs}$, $\tilde{\gamma}_{fn}^{\mu}$ and $\tilde{\gamma}_{fn}^{\nu}$ are the
anomalous dimensions of the singlets and the nonsinglets.
The $\mu$ and $\nu$ anomalous dimensions for the beam functions of the gauge bosons in the 
Laplace transform are given as
\begin{align} \label{beamanom}
&\tilde{\gamma}^{\mu}_{Ws} = 2C_A \Gamma_c \ln \frac{\mu^2}{Q_L \omega} -2\gamma_W, \ \ 
\tilde{\gamma}^{\nu}_{Ws} = 0, \nonumber \\
&\tilde{\gamma}^{\mu}_{Wn} = \tilde{\gamma}^{\mu}_{Ws}-C_A \Gamma_c \ln \frac{\mu^2}{Q_L \nu},
\ \ \tilde{\gamma}^{\nu}_{Wn} =  C_A  \Gamma_c \ln \frac{\mu}{M},
\end{align}
where the non-cusp anomalous dimension at NLO is given as $\gamma_W^{(1)} = - \beta_0 \alpha/(4\pi)$.
The cusp anomalous dimension~\cite{Korchemsky:1987wg,Korchemskaya:1992je}  $\Gamma_c(\alpha)$ 
can be expanded as
\begin{equation} 
\Gamma_c (\alpha) = \frac{\alpha}{4\pi}
\Gamma_c^0 + \Bigl( \frac{\alpha}{4\pi}\Bigr)^2 \Gamma_c^1 +\cdots, 
\end{equation} 
with
\begin{equation} 
\Gamma_c^0 =4, \ \Gamma_c^1 = \Bigl( \frac{268}{9}-\frac{4}{3}\pi^2
\Bigr) C_A -\frac{40 n_f}{9}. 
\end{equation} 
To NLL accuracy, the cusp anomalous dimension to two loops is needed.
 
The anomalous dimensions for the beam function of the gauge bosons in eq.~\eqref{beamanom}, 
the beam function of the leptons and the semi-inclusive jet functions~\cite{Chay:2021arz} can be written 
in a general form as
\begin{align}
&\tilde{\gamma}_{is}^{\mu}= 2C_i \Gamma_c \ln  \frac{\mu^2}{Q_L\omega} - 2\gamma_i,  \ \
\tilde{\gamma}_{in}^{\mu}  =  \tilde{\gamma}_{is}^{\mu} -C_A \Gamma_c \ln \frac{\mu^2}{Q_L\nu},
\nonumber \\
&\tilde{\gamma}_{is}^{\nu} =0, \ \ \tilde{\gamma}_{in}^{\nu} = C_A \Gamma_c \ln \frac{\mu}{M},
\end{align}
with $i= \ell, W$, and the group theory factors are given as $C_{\ell} = C_F$, $C_W =C_A$. 
The non-cusp anomalous dimensions at order $\alpha$ are given by
\begin{equation}
\gamma_{\ell}^{(1)} = -3C_F \frac{\alpha}{4\pi}, \ \ \gamma_W^{(1)} = -\beta_0 \frac{\alpha}{4\pi}.
\end{equation}

The evolution of the singlet collinear functions from eq.~\eqref{singev} can be written as
\begin{equation}
\tilde{f}_{is} (\mu_F) = U_{is} (\mu_F, \mu_C) \tilde{f}_{is} (\mu_C),
\end{equation}
with $i =\ell, W$, where the evolution kernel is given as
\begin{equation} \label{uis}
U_{is} (\mu_F, \mu_C) = \exp \Bigl[ 4C_i S_{\Gamma} (\mu_F, \mu_C) - 2C_i a_{\Gamma} 
(\mu_F, \mu_C) \ln \frac{\omega Q_L}{\mu_C^2} -2a_{\gamma_i} (\mu_F, \mu_C)\Bigr].
\end{equation}
The various quantities in eq.~\eqref{uis} are defined as
\begin{equation}
\int_{\mu_C}^{\mu_F} \frac{d\mu}{\mu} \Gamma_c (\alpha) \ln \frac{\mu^2}{\omega Q_L} \equiv
2S_{\Gamma} (\mu_F,\mu_C) -a_{\Gamma} (\mu_F,\mu_C) \ln \frac{\omega Q_L}{\mu_C^2}.   
 \end{equation}
where
\begin{equation}
S_{\Gamma} (\mu, \mu_i) = \int_{\alpha(\mu_i)}^{\alpha (\mu)} d\alpha 
\frac{\Gamma_c (\alpha)}{\beta(\alpha)} \int_{\alpha (\mu_i)}^{\alpha} \frac{d\alpha'}{\beta(\alpha')}, \ \ 
a_f (\mu, \mu_i) = \int_{\alpha(\mu_i)}^{\alpha (\mu)} \frac{d\alpha}{\beta(\alpha)} f(\alpha), 
\end{equation}
with $f(\alpha) = \Gamma_c (\alpha)$ or any other noncusp anomalous dimensions such as $\gamma_{\ell}$ 
or $\gamma_W$.

On the other hand, the nonsinglet collinear functions satisfy the coupled RG equations, eq.~\eqref{nonev}.
And the evolution from both RG equations are written as
\begin{equation}
\tilde{f}_{in} (\mu_F, \nu_F) = U_{in} (\mu_F, \mu_C; \nu_F) V_{in} (\nu_F, \nu_C; \mu_C) 
\tilde{f}_{in} (\mu_C, \nu_C),
\end{equation}
where the evolution kernels are given as
\begin{align}
U_{in} (\mu_F, \mu_C; \nu_F) &= U_{is} (\mu_F, \mu_C) \exp[-2C_A S_{\Gamma} (\mu_F, \mu_C)
+C_A a_{\Gamma} (\mu_F, \mu_C) \ln \frac{\nu_F Q_L}{\mu_C^2}\Bigr],
\nonumber \\
V_{in} (\nu_F, \nu_C; \mu_C) &= \exp\Bigl[ C_A a_{\Gamma} (\mu_C, M) 
\ln \frac{\nu_F}{\nu_C}\Bigr].
\end{align}

 \subsection{Hard function}
The RG equation for the hard function $\mathbf{H}_W$ in the channel $W W\rightarrow \ell_{\mu}
\overline{\ell}_{\mu}$ is written as
\begin{equation}
\frac{d}{d\ln \mu} \mathbf{H}_W = \mathbf{\Gamma}_H^W \mathbf{H}_W + 
\mathbf{H}_W \mathbf{\Gamma}_H^{W\dagger}. 
\end{equation}
The relation between the anomalous dimensions $\bm{\Gamma}_H^W$  in our basis, and  the anomalous
dimensions $\bm{\Gamma}_H^{\prime W}$ in ref.~\cite{Kelley:2010fn} is given by 
$\bm{\Gamma}_H^W =  A \bm{\Gamma}_H^{\prime W} A^{-1}$. Note that only the nondiagonal matrix
is affected by this transformation.

 The anomalous dimension matrix~\cite{Kelley:2010fn,Chiu:2009mg} can be generalized to 
include the channels $W  W \rightarrow \ell_{\mu} \overline{\ell}_{\mu}$ and 
$\ell_e \overline{\ell}_e\rightarrow \ell_{\mu} \overline{\ell}_{\mu}$. 
The anomalous dimensions $\bm{\Gamma}_i$ ($i=W, e$) of the hard 
functions $\mathbf{H}_i$ can be written as
\begin{equation} \label{gamh} 
\mathbf{\Gamma}_H^i (s,t,u)= \Bigl[ \frac{c_H}{2} \Gamma_c (\alpha ) L_i (\mu)
+\gamma_H \Bigr] \mathbf{1} + \Gamma_c (\alpha) \mathbf{M}_i (s,t,u),
\end{equation} 
where $c_H = n_{\ell} C_F+ n_W C_A$ with $n_{\ell}$ fermions and $n_W$ gauge bosons involved in the
channel. 
And $\gamma_H = n_{\ell} \gamma_{\ell}+ n_W\gamma_W$, with $L_W (\mu) = L(t)$, 
$L_e (\mu) = L(s)$. 
 The matrices $\mathbf{M}_W$ and $\mathbf{M}_e$, in the two channels, are given by 
\begin{align} \label{hardm}
\mathbf{M}_W &=
\begin{pmatrix}
\displaystyle \frac{C_A +C_F}{2}  \ln \frac{n_{12} n_{34}}{n_{13} n_{24}} & \displaystyle\frac{1}{2} 
\ln \frac{n_{13}n_{24}}{n_{14}n_{23}}& 0 \\
\displaystyle \ln \frac{n_{13}n_{24}}{n_{14} n_{23}} & \displaystyle \frac{C_A}{4}  
\ln \frac{n_{14} n_{23}}{n_{13} n_{24}} + \frac{C_F}{2}  
\ln \frac{n_{12} n_{34}}{n_{13}n_{24}}& \displaystyle\frac{-3C_A +8C_F}{4}  
\ln \frac{n_{13} n_{24}}{n_{14} n_{23}} \\
0& \displaystyle \frac{C_A}{4}  \ln \frac{n_{13} n_{24}}{n_{14} n_{23}} & \displaystyle 
\frac{C_A}{4} \ln \frac{n_{14} n_{23}}{n_{13} n_{24}} +\frac{C_F}{2}  \ln \frac{n_{12} n_{34}}{n_{13} n_{24}}
\end{pmatrix}, \nonumber \\
\mathbf{M}_e &= \begin{pmatrix} 
\displaystyle \Bigl(2C_F - C_A/2\Bigr) \ln \frac{n_{13}
n_{24}}{n_{14} n_{23}} -\frac{C_A}{2} \ln \frac{n_{12} n_{34}}{n_{14} n_{23}} &&
\displaystyle \ln \frac{n_{13} n_{24}}{n_{14} n_{23}} \\ \displaystyle -C_F \Bigl(C_F
-C_A/2\Bigr) \ln \frac{n_{13} n_{24}}{n_{14} n_{23}} & &0 
\end{pmatrix}.
\end{align}
 
The evolution of the hard functions from the hard scale $\mu_H$ to the factorization scale $\mu_F$ 
is written in the form
\begin{equation}
\mathbf{H}_i (\mu_F)  = \Pi_H^i (\mu_F,\mu_H) \bm{\Pi}_H^i (\mu_F,\mu_H)  \mathbf{H}_i (\mu_H)
\bm{\Pi}_H^{i\dagger} (\mu_F,\mu_H),
\end{equation}
where $\Pi_H^i (\mu_F,\mu_H)$ describes the evolution from the anomalous dimension, 
proportional to the idendity matrix, while $\bm{\Pi}_H^{i} (\mu_F,\mu_H)$ 
is the evolution kernel from $\mathbf{M}_i$.
They are given as
\begin{align} \label{hardevol}
\Pi_H^i (\mu_F, \mu_H) &= \exp\Bigl[ -2c_H S(\mu_F, \mu_H) + c_H a_{\Gamma}
(\mu_F,\mu_H) L_i (\mu_H)  + 2  a_{\gamma_H} (\mu_F,\mu_H)   \Bigr], \nonumber \\
\bm{\Pi}_H^i (\mu_F,\mu_H) &= \exp \Bigl[ a_{\Gamma} (\mu_F,\mu_H) \mathbf{M}_i\Bigr].
\end{align}
 
\subsection{Soft function}
The RG equation for the soft function $\tilde{\mathbf{S}}$ is written as
\begin{equation} \label{rgsoft}
\frac{d}{d\ln \mu} \tilde{\mathbf{S}} = \tilde{\bm{\Gamma}}_S^{\mu\dagger} \tilde{\mathbf{S}}
+  \tilde{\mathbf{S}}   \tilde{\bm{\Gamma}}_S^{\mu},
\end{equation}
where $\tilde{\bm{\Gamma}}_S^{\mu}$ is the $\mu$-anomalous dimension matrix.
The relation of the soft functions between our basis $T_I$ and another basis $T'_I$ is obtained by 
requiring that $\mathrm{tr} (\mathbf{H}'\mathbf{S}')=  \mathrm{tr} (\mathbf{H} \mathbf{S})$,
from which we obtain that $\mathbf{S} = (\mathbf{A}^{-1})^{\dagger} \mathbf{S}' \mathbf{A}^{-1}$, and
$\tilde{\bm{\Gamma}}_S^{\mu} = \mathbf{A}\tilde{\bm{\Gamma}}_S^{\prime\mu} \mathbf{A}^{-1}$.

At order $\alpha$, eq.~\eqref{rgsoft} is written as
\begin{equation} \label{rgsoft1}
\frac{d}{d\ln \mu} \tilde{\mathbf{S}}^{(1)} =  \tilde{\bm{\Gamma}}_S^{\mu\dagger (1)} 
\tilde{\mathbf{S}}^{(0)} + \tilde{\mathbf{S}}^{(0)} \tilde{\bm{\Gamma}}_S^{\mu (1)},
\end{equation}
where $\tilde{\mathbf{S}}^{(1)}$ is the renormalized soft function. 
The $\mu$-soft anomalous dimensions can be extracted from the requirement that the sum of the anomalous 
dimensions from all the factorized parts should cancel.  The $\mu$-soft anomalous dimensions at NLO 
are written as
\begin{equation}
\tilde{\bm{\Gamma}}_{iS}^{\mu(1)} = -\Bigl(\tilde{\bm{\Gamma}}_{iH}^{(1)}  + 
(\tilde{\gamma}_i^{\mu (1)}  +\tilde{\gamma}_{J}^{\mu (1)}  ) 
\otimes \mathbf{1}\Bigr),
\end{equation}
where $\tilde{\gamma}_i^{\mu}$ are the anomalous dimensions of the beam functions ($i=\ell, W$),
and $\tilde{\gamma}_{J}^{\mu}$ is that of the jet function.
And the RG equation with respect to the rapidity scale $\nu$ is written as
\begin{equation}
\frac{d}{d\ln \nu} \tilde{\mathbf{S}} = \tilde{\Gamma}_S^{\nu} \tilde{\mathbf{S}}.
\end{equation}
 
The soft anomalous dimensions for $W W\rightarrow \ell_{\mu} \overline{\ell}_{\mu}$
 are given by
\begin{align} \label{softm}
\Bigl(\Gamma_S^{k\mu} \Bigr)_W &= \Gamma_c  \Bigl[ 
\frac{c_H}{4}\ln \frac{Q_L^4}{n_{13} n_{24} \mu^4}
+\frac{1}{2}(C_F-C_A) \ln \frac{n_{12}}{n_{34}}  \Bigr]
 -\Gamma_c \mathbf{M}_W 
+\frac{kC_A}{2} \Gamma_c \ln \frac{\mu^2}{Q_L \nu}, \nonumber \\
\Bigl(\Gamma_S^{k\nu}\Bigr)_W &= -kC_A \Gamma_c \ln \frac{\mu}{M},
\end{align}
where $k$  ($k= 0, 2, 3, 4$)  is the number of the nonsinglets involved in the process.\footnote{The case with
$k=1$ vanishes due to the color conservation, and only the even number of nonsinglets is allowed in SU(2).}
We can write the soft anomalous dimensions in a compact form including both channels
$\ell_e \overline{\ell}_e \rightarrow \ell_{\mu} \overline{\ell}_{\mu}$ and $W W \rightarrow 
\ell_{\mu} \overline{\ell}_{\mu}$, respectively. The Laplace transformed $\mu$ and $\nu$ 
soft anomalous dimensions can be written as
\begin{align} \label{totsoft}
\Bigl(\tilde{\Gamma}_{Sk}^{\mu}\Bigr)_i &= \Gamma_c   
\Bigl[\frac{c_H}{4} \ln \frac{Q_L^4}{\mu^4 N_i} +B_i\Bigr] \mathbf{1} -\Gamma_c  \mathbf{M}_i
+k\frac{C_A}{2} \Gamma_c \ln \frac{\mu^2}{Q_L \nu} \mathbf{1},  \nonumber \\
\Bigl(\tilde{\Gamma}_{Sk}^{\nu}\Bigr)_i &= -k C_A \Gamma_c \ln \frac{\mu}{M},
\end{align}
where $i= \ell$ or $W$. The quantities in eq.~\eqref{totsoft} are given as
\begin{equation}
N_{\ell} = n_{12} n_{34}, \ \ N_W = n_{13}n_{24},  \ \ 
 B_{\ell} =0,  \ \ B_W = \frac{1}{2}(C_F -C_A) \ln \frac{n_{12}}{n_{34}}.
\end{equation}
Note that  $B_W=0$ for $n_{12}= n_{34}$ when the jets are back-to-back. 
The list of the anomalous dimensions for all the factorized parts  is shown
in table~\ref{anomtab}.

\begin{table}[t] 
\begin{center}
\begin{tabular}{cccc} \hline
Functions&Type &$\mu$-anomalous dimensions & $\nu$-anomalous dimensions\\  \hline
Hard function &$-$ &$\bm{\Gamma}_H^i = \Bigl[\frac{c_H}{2} \Gamma_c L_i 
+\gamma_H\Bigr] \mathbf{1} + \Gamma_c \mathbf{M}_i$ & 0\\
\multirow{2}{*}{Lepton Beam}& singlet& $\tilde{\gamma}_{\ell s}^{\mu} = 2C_F \Gamma_c \ln 
  \frac{\mu^2}{Q_L \omega}  -2\gamma_{\ell}$ & 0\\
&nonsinglet& $\tilde{\gamma}_{\ell n}^{\mu} = \tilde{\gamma}_{\ell s}^{\mu} -C_A \Gamma_c \ln  
\frac{\mu^2}{Q_L \nu}$ & $\tilde{\gamma}_{\ell n}^{\nu} = C_A \Gamma_c \ln \frac{\mu}{M}$ \\
\multirow{2}{*}{Boson Beam}& singlet& $\tilde{\gamma}_{Ws}^{\mu} = 2C_A \Gamma_c \ln   
\frac{\mu^2}{Q_L \omega} -2\gamma_W$ & 0\\
&nonsinglet& $\tilde{\gamma}_{Wn}^{\mu} =\tilde{\gamma}_{Ws}^{\mu} -C_A \Gamma_c \ln  
\frac{\mu^2}{Q_L \nu}$ & $\tilde{\gamma}_{bn}^{\nu} = C_A \Gamma_c \ln \frac{\mu}{M}$\\
\multirow{2}{*}{Muon jet}& singlet& $\tilde{\gamma}_{\ell s}^{\mu} = 2C_F \Gamma_c \ln 
 \frac{\mu^2}{Q_L \omega}  -2\gamma_{\ell}$ & 0\\
&nonsinglet& $\tilde{\gamma}_{\ell n}^{\mu} = \tilde{\gamma}_{\ell s}^{\mu} -C_A \Gamma_c \ln  
\frac{\mu^2}{Q_L \nu}$ & $\tilde{\gamma}_{\ell n}^{\nu} = C_A \Gamma_c \ln \frac{\mu}{M}$ \\
\multirow{2}{*}{Soft function}&singlet &$(\tilde{\bm{\Gamma}}_S^{\mu})_i 
=\Gamma_c \Bigl[ \frac{c_H}{4} \ln
\frac{Q_L^4}{\mu^4 N_i} +B_i\Bigr] \mathbf{1} -\Gamma_c \mathbf{M}_i$ & 0\\
&nonsinglet& $(\tilde{\bm{\Gamma}}_{Sk}^{\mu})_i =(\bm{\Gamma}_S^{\mu})_i 
+k\frac{C_A}{2} \Gamma_c \ln 
\frac{\mu^2}{Q_L \nu} \mathbf{1}$& $(\tilde{\Gamma}_{Sk}^{\nu})_i = -kC_A\Gamma_c 
\ln \frac{\mu}{M}$ \\
\hline
\end{tabular}
\caption{Anomalous dimensions of the factorized parts.\label{anomtab}} 
\end{center}
\end{table}

The evolution of the Laplace-transformed soft functions is written as
\begin{equation}
\tilde{S}_k^i (\mu_F, \nu_F) = \Pi_{Sk}^i (\mu_F, \mu_S; \nu_F, \nu_S) \bm{\Pi}^{i\dagger} 
(\mu_F, \mu_S) \tilde{S}_k^i (\mu_S, \nu_S) \bm{\Pi}_S^i (\mu_F, \mu_S),
\end{equation}
where $\bm{\Pi}^i$ is the evolution kernel from $\mathbf{M}_i$  ($i=\ell, W$). And  $\Pi_{Sk}^i$ is 
the kernel from the diagonal part, with $k$ nonsinglets.  $\Pi_{Sk}^i$ involves a 
double evolution with respect to $\mu$ and $\nu$. Since the order of the evolution is irrelevant, we first 
evolve with respect to $\nu$, and then  $\mu$. As a result, $\Pi_{Sk}^i$ can be written as
\begin{equation}
\Pi_{Sk}^i (\mu_F, \mu_S, \nu_F, \nu_S) = U_{Sk}^i (\mu_F,\mu_S; \nu_F) 
V_{Sk}^i (\nu_F, \nu_S; \mu_S),
\end{equation}
where
\begin{align}
V_{Sk}^i (\nu_F, \nu_S; \mu_S) &= \exp \Bigl[ -k C_A a_{\Gamma} (\mu_S, M) 
\ln \frac{\nu_F}{\nu_S}\Bigr],
 \\
U_{Sk}^i (\mu_F,\mu_S; \nu_F)  &= \exp \Bigl[ -2c_H S_{\Gamma} (\mu_F, \mu_S)  
+\frac{c_H}{2}a_{\Gamma} (\mu_F, \mu_S) \ln \frac{Q_L^4}{\mu_S^4 N_i}
\nonumber \\ 
&+  B_i a_{\Gamma} (\mu_F, \mu_S)  
  +2kC_A S_{\Gamma} (\mu_F,\mu_S)+kC_A a_{\Gamma} (\mu_F, \mu_S) 
\ln \frac{\mu_S^2}{Q_L \nu_F}\Bigr].  \nonumber
\end{align}
 
\section{Numerical estimates of the 2-jettiness in SU(2) near   threshold \label{numeric}}
In order to estimate the singlet and the nonsinglet contributions to the 2-jettiness, we confine
ourselves to the threshold limit in which the two jets are emitted back-to-back near $\theta = \pi/2$ 
(perpendicular to the beam axis) at NLL.
It not only simplifies the kinematics, but also is available from the current information 
obtained so far. We explain these points first, and present the numerical analysis accordingly.
 
Firstly, the anomalous dimensions depend on the angles among the light-like directions. [See,
for example, eqs.~\eqref{hardm} and \eqref{softm}.] Therefore 
the evolution of the 2-jettiness in eq~\eqref{twojet} involves the integration over these angles. 
The numerical integration over the angles can be performed, but we would rather consider the differential 2-jettiness 
$d\sigma/d\mathcal{T}_2 dt$ at $\theta=\pi/2$. It corresponds to the back-to-back jets, perpendicular
to the beam directions. 

Secondly, the information on the PDFs is not available, in contrast to QCD. That is, the information of 
the probability in finding a lepton with a certain fraction of the longitudinal momentum is lacking. 
 We naively assume that the beam functions have the form
$B_i (t,x, M, \mu_C) =\delta (t) \delta (1-x)$  for $i=\ell, W$ at tree level, which amounts to the PDFs 
with the form  $f_i  (x)\propto\delta (1-x)$ at $\mu_C$. Actually the detailed functional forms at 
tree level and even at higher orders 
are not known, hence it should be understood that we pick up the contribution from the region $x=1$.
These delta functions will be adopted in our
analysis. If we consider the normalized 2-jettiness  $(d\sigma_i/d\mathcal{T}_2 dt)/(d\sigma_i^{(0)}
/d\mathcal{T}_2 dt)$, where $d\sigma_i^{(0)}/d\mathcal{T}_2 dt$ is the Born
result, the uncertainties on the PDF or the beam functions can be somewhat reduced in the ratio. 

Finally, we have computed the hemisphere soft functions only at NLO, from which the anomalous dimensions
can be obtained. The non-hemisphere part of the soft function is finite, and is needed for the computation
beyond NLL. Here we confine to NLL accuracy such that only  the hemisphere 
soft function is needed. 

In summary, we consider the 2-jettiness near threshold  and at NLL accuracy,  and  the jets are 
produced  perpendicular to the beam directions.
In this case, $\omega_1 =\omega_2 = \omega_3 = \omega_4 =\omega$, $t= (p_1 - p_3)^2  = -\omega_1
\omega_3 n_1 \cdot n_3/2 = -\omega^2 (1-\cos\theta )/2$ by choosing $n_1^{\mu} = (1,0, 0,1)$,
$n_2^{\mu} = (1, 0, 0, -1)$, $n_3^{\mu} = (1, \sin\theta , 0, \cos\theta )$ and
$n_4^{\mu} = (1, -\sin\theta , 0, -\cos\theta )$.  

With the SU(2) gauge interaction, the nonsinglet beam functions for the gauge bosons do 
not contribute ($d^{abc}=0$), and the beam functions for the leptons and the semi-inclusive muon jet functions
are employed at NLL.
The product of the hard and soft functions for $WW\rightarrow \ell_{\mu} \overline{\ell}_{\mu}$ is written as
\begin{align}
&\sum_{IJ} H_{IJ}(\mu_F)\tilde{S}_{JI}^{00gh} (\mu_F, \nu_F) \nonumber \\
&= \exp \Bigl[ -4(C_A+ C_F) S_{\gamma}
(\mu_F, \mu_H) -2 (C_A +C_F) a_{\Gamma} (\mu_F, \mu_H) \ln \frac{\mu_H^2}{\omega^2 n_{13}}
\nonumber \\
&+4\Bigl( a_{\gamma_W}(\mu_F, \mu_H) +a_{\gamma_{\mu}} (\mu_F, \mu_H) \Bigr) \Bigr] \nonumber \\
&\times \exp \Bigl[ -4(C_A +C_F) S_{\Gamma} (\mu_F, \mu_S) + 2(C_A+C_F) a_{\Gamma}
(\mu_F, \mu_S) \ln \frac{Q_L^2}{\mu_S^2 n_{13}}\Bigr] \nonumber \\
&\times \exp \Bigl[ k\Bigl( 2C_A S_{\Gamma} (\mu_F, \mu_S) + C_A a_{\Gamma} (\mu_F, \mu_S)
\ln \frac{\mu_S^2}{Q_L \nu_F} -C_A a_{\Gamma} (\mu_S, M) \ln \frac{\nu_F}{\nu_S} \Bigl) \Bigr] 
\nonumber  \\
&\times \mathrm{Tr} \Bigl( H^{(0)} \exp \Bigl[ (a_{\Gamma} (\mu_S, \mu_H) \mathbf{M}^{\dagger}\Bigr]
\mathbf{S}^{(0)k} \exp \Bigl[ a_{\Gamma} (\mu_S, \mu_H) \mathbf{M}\Bigr] \Bigr).
\end{align}
Note that there should be an even number of nonsinglets in the jet functions.

Note that the inverse Laplace transform of the collinear and soft functions is given by
\begin{equation}
\mathcal{L}^{-1} [Q_L^a] = \mathcal{L}^{-1} \Bigl[ \Bigl(\frac{1}{s\exp (\gamma_{\mathrm{E}}) } \Bigr)^{a}\Bigr]=\frac{1}{\mathcal{T}_2 \Gamma(a)} \Bigl(\frac{\mathcal{T}_2}{ 
\exp (\gamma_{\mathrm{E}})}\Bigr)^a = \frac{\exp \Bigl(a\ln \mathcal{T}_2 \Bigr)}{\mathcal{T}_2
\Gamma(a) \exp (a\gamma_{\mathrm{E}})},
\end{equation}
which is also valid for negative $a$~\cite{Becher:2006mr,Ligeti:2008ac}. Therefore
the differential 2-jettiness from $W W \rightarrow \mu^- \mu^+$ is written as
\begin{align} \label{ww2jet}
\frac{d\sigma_W}{d\mathcal{T}_2 dt} &=  \frac{4\pi\alpha^2}{\omega^4} \sum_{k=0,2}  
\Bigl(\frac{1}{6}\Bigr)^2 \Bigl( \frac{1}{4}\Bigr)^{2-k} \Bigl(\frac{1}{2}\Bigr)^k
\frac{1}{\mathcal{T}_2\Gamma (a_k) \exp [a_k\gamma_{\mathrm{E}}]} \nonumber  \\
& \times \exp \Bigl[ -4(C_A+ C_F) S_{\Gamma}
(\mu_F, \mu_H) -2 (C_A +C_F) a_{\Gamma} (\mu_F, \mu_H) \ln \frac{\mu_H^2}{\omega^2 n_{13}}
\nonumber  \\
 &+4\Bigl( a_{\gamma_W}(\mu_F, \mu_H) +a_{\gamma_{\ell}} (\mu_F, \mu_H) \Bigr) \Bigr]  
\nonumber \\
&\times \exp \Bigl[ 8(C_A +C_F) S_{\Gamma} (\mu_F, \mu_C) -4(C_A+C_F) a_{\Gamma} 
(\mu_F, \mu_C) \ln \frac{\omega \mathcal{T}_2}{\mu_C^2} \nonumber \\
&-4\Bigl( a_{\gamma_W} (\mu_F, \mu_C) + a_{\gamma_{\ell}} (\mu_F, \mu_C) \Bigr) \nonumber \\
&+ kC_A\Bigl( -2 S_{\Gamma} (\mu_F, \mu_C) + a_{\Gamma} (\mu_F, \mu_C)
\ln \frac{\nu_F \mathcal{T}_2}{\mu_C^2} +a_{\Gamma} (\mu_F, M) \ln \frac{\nu_F}{\nu_C}\Bigr) 
\Bigr] \nonumber \\
&\times \exp \Bigl[ -4 (C_ A +C_F) S_{\Gamma} (\mu_F, \mu_S) +2(C_A+C_F) a_{\Gamma} 
(\mu_F, \mu_S) \ln \frac{\mathcal{T}_2^2}{\mu_S^2 n_{13}} \nonumber \\
&+ kC_A\Bigl( 2S_{\Gamma} (\mu_F, \mu_S) + a_{\Gamma} (\mu_F, \mu_S) 
\ln \frac{\mu_S^2} {\nu_F \mathcal{T}_2} -a_{\Gamma} (\mu_S, M) \ln \frac{\nu_F}{\nu_S}\Bigr)
\Bigr] \nonumber \\
&\times \mathrm{Tr} \Bigl\{ \mathbf{H}_W^{(0)} \exp \Bigl[ a_{\Gamma} (\mu_S, \mu_H) 
\mathbf{M}_W^{\dagger}\Bigr] \mathbf{S}_W^{(0)k} \exp\Bigl[ a_{\Gamma} (\mu_S, \mu_H) 
\mathbf{M}_W^{\dagger}\Bigr] \Bigr\},
\end{align}
where $a_k$ is given by
\begin{equation}
a_k = [4(C_A+C_F)-kC_A] a_{\Gamma} (\mu_C, \mu_S),
\end{equation}
and $n_{13} = (1-\cos\theta)/2$. And the numerical factors in front come from the product of the
color factors $k_i^a$ in eq.~\eqref{twojet} and the muon projections in eq.~\eqref{jetex}.

The matrix $\mathbf{M}_W$ near threshold is given as
\begin{equation}
\mathbf{M}_W^{\mathrm{th}} = \begin{pmatrix} 
 -(C_F +C_A) \ln  n_{13} & \displaystyle \ln \frac{n_{13} }{1-n_{13}}& 0 \\ 
 \displaystyle 2\ln \frac{n_{13}}{1-n_{13}} & \displaystyle \frac{C_A}{2}
\ln \frac{1- n_{13}}{n_{13}} - C_F \ln n_{13}&
\displaystyle \frac{1}{2} (-3C_A + 8C_F) \ln \frac{n_{13}}{1- n_{13}} \\
0& \displaystyle \frac{C_A}{2} \ln \frac{n_{13} }{1- n_{13}} &
\displaystyle \frac{C_A}{2}
\ln \frac{1- n_{13}}{n_{13}} +  C_F  \ln  n_{13}  
\end{pmatrix}. 
\end{equation}
The hard functions and the soft functions at LO are given as
\begin{align}
\mathbf{H}_W^{(0)} &= \frac{1}{n_{13}(1-n_{13})}\begin{pmatrix}
\displaystyle \frac{1-2n_{13}(1-n_{13})}{N^2} & \displaystyle\frac{(1-n_{13})^4-n_{13}^4}{N}  &
\displaystyle \frac{1-2n_{13}(1-n_{13})}{N}  \\
\displaystyle\frac{(1-n_{13})^4-n_{13}^4}{N}  &  (1-2n_{13})^2\Bigl( 1-2n_{13}(1-n_{13})\Bigr) & 
(1-n_{13})^4-n_{13}^4\\
\displaystyle \frac{1-2n_{13}(1-n_{13})}{N} &(1-n_{13})^4-n_{13}^4&1-2n_{13}(1-n_{13})
\end{pmatrix}, \nonumber \\
 \mathbf{S}_W^{(0)0} &= S^{(0)} (0, 0, 0, 0)  =
\begin{pmatrix}
2C_F C_A^2&0&0 \\
0&C_F C_A^2 &0 \\
0&0&\displaystyle 0
\end{pmatrix}, \nonumber \\
\mathbf{S}_W^{(0)2} &= S^{(0)} (0, 0, 1, 1) =
\begin{pmatrix}
2C_F C_A&0&0 \\
0&\displaystyle C_A (C_F -C_A/2) &0 \\
0&0& 0
\end{pmatrix} (34).
\end{align} 
Here $(a_1 a_2 \cdots a_n)$ denotes $\mathrm{Tr} (t^{a_1} t^{a_2} \cdots t^{a_n})$. All the other 
soft matrices at treel level  are presented in appendix~\ref{softzm}.

Note that the hard and soft anomalous dimensions, hence the corresponding evolution kernels, depend
on $n_{13} = (1-\cos\theta)/2$, as well as the hard functions themselves. And
we fix $\theta = \pi/2$ in the numerical estimation. It corresponds to the back-to-back jets perpendicular
to the beam axis. 
 
Similarly, we can write the differential 2-jettiness for the channel $\ell_e \overline{\ell}_e\rightarrow 
\ell_{\mu} \overline{\ell}_{\mu}$. In contrast to the channel 
$W W\rightarrow \ell_{\mu} \overline{\ell}_{\mu}$, there are nonsinglet contributions from the electron beam
functions. The only constraint is that there should be even number of nonsinglet contributions as a whole. 
It is given as
\begin{align}
\frac{d\sigma_e}{d\mathcal{T}_2 dt} &= \frac{64\pi\alpha^2}{\omega^4} 
\sum_{k=0,2,4} \Bigl(\frac{1}{4}\Bigr)^{4-k} \Bigl( \frac{1}{2}\Bigr)^k
\frac{1}{\mathcal{T}_2\Gamma (a_k) \exp (a_k\gamma_{\mathrm{E}})} \nonumber \\
&\times \exp \Bigl[ -8C_F S_{\Gamma} (\mu_F, \mu_H) -4C_F a_{\Gamma} (\mu_F, \mu_H) 
\ln \frac{\mu_H^2}{\omega^2} + 8 a_{\gamma_{\ell}} (\mu_F, \mu_H)  \Bigr] \nonumber \\
&\times \exp\Bigl[ 16 C_F S_{\Gamma} (\mu_F, \mu_C)  -8C_F a_{\Gamma} (\mu_F, \mu_C)
\ln \frac{\omega \mathcal{T}_2}{\mu_C^2} -8 a_{\gamma_{\ell}} (\mu_F, \mu_C) \nonumber \\
&+ k C_A\Bigl( -2 S_{\Gamma} (\mu_F, \mu_C) +  a_{\Gamma} (\mu_F, \mu_C) \ln
\frac{\nu_F \mathcal{T}_2}{\mu_C^2}  + a_{\Gamma} (\mu_F, M) \ln \frac{\nu_F}{\nu_C} 
\Bigl) \Bigr] \nonumber \\
&\times \exp \Bigl[ -8 C_F S_{\Gamma} (\mu_F, \mu_S)  +4C_F a_{\Gamma}  (\mu_F, \mu_S) 
\ln \frac{\mathcal{T}_2^2}{\mu_S^2} \nonumber 
\end{align}
\begin{align}
&+kC_A \Bigl( 2  S_{\Gamma}  (\mu_F, \mu_S) + a_{\Gamma}  (\mu_F, \mu_S) 
\ln \frac{\mu_S^2}{\nu_F \mathcal{T}_2} - a_{\Gamma}  (\mu_F, M)  \ln \frac{\nu_F}{\nu_S} \Bigr)
\Bigr] \nonumber \\
&\times \mathrm{Tr} \Bigl\{ \mathbf{H}_e^{(0)} \exp \Bigl[ a_{\Gamma} (\mu_S, \mu_H)
\mathbf{M}_e^{\dagger}\Bigr] \mathbf{S}_e^{(0) k} \exp \Bigl[  a_{\Gamma} (\mu_S, \mu_H)
\mathbf{M}_e \Bigr] \Bigr\},
\end{align}
where we choose an even number of nonsinglets out of $T^e$, $T^f$, $T_g$ and $T^h$, which we
denote as $k$. In this case, $a_k$ is given as
\begin{equation} \label{akee}
a_k = (8C_F-kC_A) a_{\Gamma} (\mu_C, \mu_S).
\end{equation}
The numerical factor in front comes from the product of the
color factors $k_i^a$ in eq.~\eqref{twojet} and the muon projections in eq.~\eqref{jetex}.

Here all the matrices $\mathbf{M}_e$, $\mathbf{H}_e^{(0)}$ and $\mathbf{S}^{(0)}$ are 
$2\times 2$ matrices. The matrix $\mathbf{M}_e$ at threshold is written as
\begin{equation} \label{me} 
\mathbf{M}_e =  \begin{pmatrix} 
\displaystyle \Bigl(4C_F - C_A \Bigr) \ln \frac{n_{13}}
{1- n_{13}} +C_A \ln (1-n_{13}) &&
\displaystyle 2 \ln \frac{n_{13}}{1-n_{13}} \\ \displaystyle -C_F \Bigl(2C_F
-C_A\Bigr) \ln \frac{n_{13} }{1-n_{13}} & &0 
\end{pmatrix}.
 \end{equation} 
The hard function $\mathbf{H}_e^{(0)}$ is given as
\begin{equation}
\mathbf{H}_e^{(0)} = \begin{pmatrix}
 (1-n_{13})^2  & 0 \\
0&0
\end{pmatrix}, 
\end{equation}
and the soft functions can be found in ref.~\cite{Chay:2021arz}.
The tree-level soft factors are given as follows:  For the singlet with $k=0$, 
\begin{equation}
\label{s0000}
\mathbf{S}_e^{(0)} (0,0,0,0) = \begin{pmatrix}
C_A C_F /2 & 0 \\
0& C_A^2
\end{pmatrix}, 
\end{equation}
and there are six possible nonsinglet soft functions with $k=2$ as
\begin{align} \label{s1100}
\mathbf{S}_e^{(0)} (1,1,0,0) &=  \frac{1}{2}\begin{pmatrix}
\displaystyle C_F - \frac{C_A}{2}   & 0 \\
0& 2 C_A 
\end{pmatrix}(12),   \  \mathbf{S}_e^{(0)} (0,0,1,1) =  \frac{1}{2} \begin{pmatrix}
\displaystyle C_F - \frac{C_A}{2}   & 0 \\
0& 2C_A
\end{pmatrix}(34), \nonumber \\
\mathbf{S}_e^{(0)} (1,0,1,0) & =  \frac{1}{2} \begin{pmatrix}
\displaystyle 2C_F - \frac{C_A}{2}   & 1 \\
1& 0
\end{pmatrix}(13), \ \mathbf{S}_e^{(0)} (0,1,0,1) =  \frac{1}{2}\begin{pmatrix}
\displaystyle 2C_F - \frac{C_A}{2}    &1 \\
1& 0 
\end{pmatrix}(24),    \nonumber \\
\mathbf{S}_e^{(0)} (1,0,0,1) &=  \frac{1}{2}\begin{pmatrix}
 2C_F - C_A    & 1 \\
1& 0 
\end{pmatrix}(14),   \  \mathbf{S}_e^{(0)} (0,1,1,0) =  \begin{pmatrix}
\displaystyle C_F - \frac{C_A}{2}   &\displaystyle \frac{1}{2} \\
\displaystyle\frac{1}{2}& 0
\end{pmatrix}(23).
\end{align}
Finally the soft function with  $k=4$ is given by
 \begin{align} \label{s1111}
\mathbf{S}_e^{(0)} (1,1,1,1) &=  \frac{1}{2} \begin{pmatrix}
\displaystyle  C_F - \frac{C_A}{2}  & 0  \\
1    &0
\end{pmatrix} (1243) + \frac{1}{2}
\begin{pmatrix}
\displaystyle C_F -\frac{C_A}{2}  & 1  \\
0  &0
\end{pmatrix} (1342) \nonumber \\
&+ \begin{pmatrix}
 \displaystyle (C_F - \frac{C_A}{2})^2   &  \displaystyle C_F - \frac{C_A}{2} \\
 \displaystyle C_F - \frac{C_A}{2}   & 1
\end{pmatrix} (12)(34) +
\begin{pmatrix}
\displaystyle\frac{1}{4}   & 0  \\
0  &0
\end{pmatrix} (13)(24).
\end{align}
\begin{figure}[h] 
 \centering
 \begin{subfigure}[h]{\textwidth}
 \centering
\includegraphics[height=8. cm]{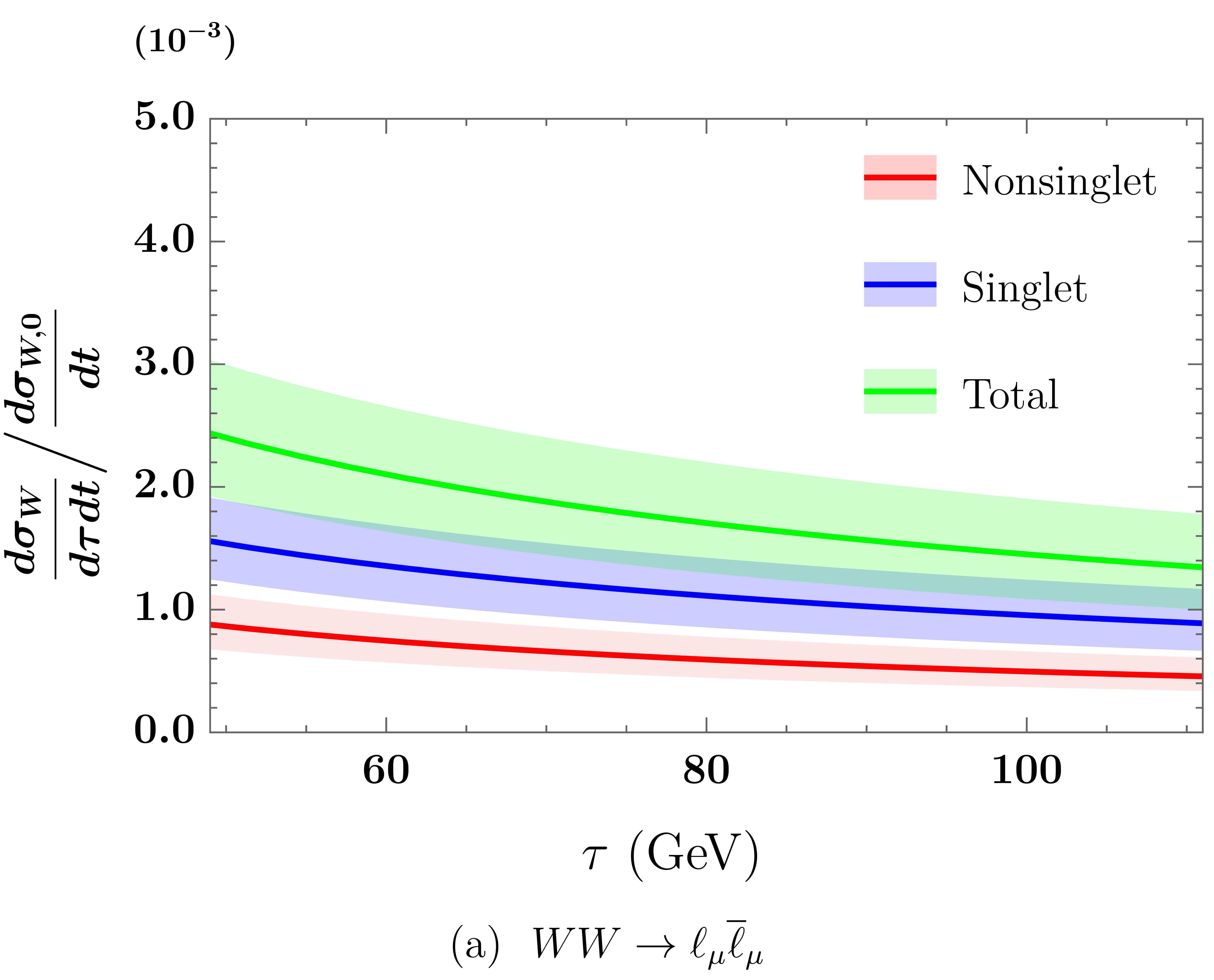} 
 \label{wwjet}
\end{subfigure}
\hfill
 \centering
 \begin{subfigure}[h]{ \textwidth}
 \centering
\includegraphics[height=8. cm]{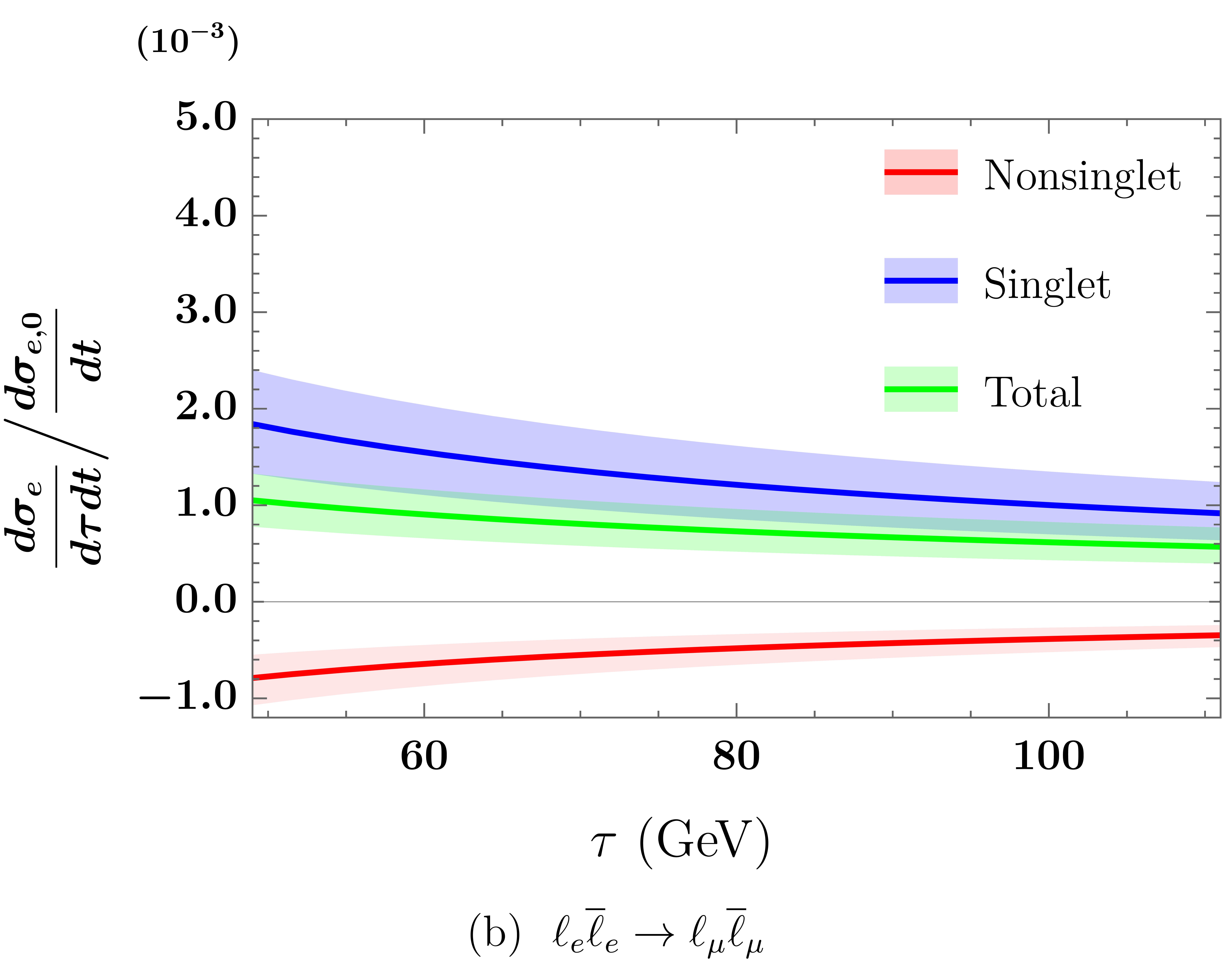} 
 \label{eejet}
\end{subfigure}\vspace{-0.1cm}
\caption{\label{jetnum}\baselineskip 3.0ex  The contributions to the 2-jettiness at NLL from the  singlet (blue),  
the nonsinglet (red), and the total (green) contributions in (a) 
$WW\rightarrow \ell_{\mu} \overline{\ell}_{\mu}$, 
(b) $\ell_e \overline{\ell}_e \rightarrow \mu^- \mu^+$ near threshold. The bands show the theoretical 
uncertainties.} 
\end{figure}

The plots of the 2-jettiness near threshold at $\theta = \pi/2$ with 50 GeV $\leq \mathcal{T} \leq$ 110 GeV
are shown in fig.~\ref{jetnum}. 
The singlet, the  nonsinglet, and the total contributions are shown respectively with the red, 
blue, and green curves with the bands showing the theoretical uncertainties at NLL. By
normalizing with the tree-level quantities, the uncertainties from the beam functions for 
the gauge bosons and the electrons in the initial state may be alleviated because there is a partial 
cancellation at higher orders.  

In fig.~\ref{jetnum} (a), the contribution to the jettiness from $W W \rightarrow 
\ell_{\mu} \overline{\ell}_{\mu}$ is shown, and the nonsinglet contribution is about 50 \% of the singlet 
contribution. It is expected because the nonsinglet contributions are more suppressed due to the additional 
evolution with respect to $\nu$.  In fig.~\ref{jetnum} (b), the corresponding jettines from 
$\ell_e \overline{\ell}_e \rightarrow \ell_{\mu} \overline{\ell}_{\mu}$ is shown. 
It may look surprising that the nonsinglet contribution from 
$\ell_e \overline{\ell}_e \rightarrow \ell_{\mu} \overline{\ell}_{\mu}$ becomes negative. The origin comes
from eq.~\eqref{akee}. When $k=4$, $a_4$ becomes negative due to the different group theory factors, 
and the contribution from $k=4$ is dominant, 
which causes the jettiness from the nonsinglets to be negative. However, it is not a worrisome problem 
because the nonsinglet contribution alone is not physical, but the total sum with the singlet contribution is physical.
The physical jettiness, that is, the sum of the singlet and the nonsinglet contributions  is positive in both cases. 
Still, the magnitude of the nonsinglet contribution
is about 50 \% of the singlet contribution. Therefore the nonsinglet contribution in $W W \rightarrow 
\ell_{\mu} \overline{\ell}_{\mu}$ is enhanced about 1.5 times compared to the singlet contribution only, while
the nonsinglet contribution in $\ell_e \overline{\ell}_e \rightarrow \ell_{\mu} \overline{\ell}_{\mu}$ 
is diminished about by half due to the negative nonsinglet contribution.

Since we do not have information on the beam functions beyond the required accuracy, we do not try 
to estimate the
total contribution to the 2-jettiness by adding the contributions from both channels. However, it is clear 
that the nonsinglet
contributions contribute to the 2-jettiness in the weak interaction significantly in both channels, 
compared to QCD, in which there is a singlet contribution only. 

\section{Conclusions\label{conc}}
We have analyzed the 2-jettiness in high-energy electroweak scattering in which muon and anti-muon dijets 
are observed. The underlying processes consist of $\ell_e \overline{\ell}_e , WW \rightarrow 
\ell_{\mu} \overline{\ell}_{\mu}$. 
The central issue  in considering the 2-jettiness in weak interaction  stems from 
the main difference between the electroweak and the QCD processes. That is, there exist 
gauge nonsinglets (electrons, muons or neutrinos) in weak interaction, while only the color singlets, or the 
hadrons are observed in strong interaction. This difference results in 
many interesting aspects, which do not appear in QCD, and exhibits more intricate structure of the 
factorization.

In inclusive quantities in QCD, the Sudakov logarithm from virtual corrections is cancelled by
that from real contributions.  For this reason, the PDF is free of the Sudakov logarithm and it satisfies the 
DGLAP evolution equation.  However, if we consider 
exclusive processes, the phase space in the virtual contribution does not coincide with that in the real 
contribution. Therefore the residual Sudakov logarithm survives. As a result, the beam
function with the small lightcone momentum induces the Sudakov logarithm and obeys a different 
evolution equation. This type of the Sudakov logarithm appears both in QCD and in weak interaction.

In weak interaction, additional Sudakov logarithms appear even in inclusive quantities because the observed
particles include gauge nonsinglets. In this case, the group theory factors in the virtual
contributions and in the real contributions are different. Therefore even for inclusive quantities, the sum
of the virtual and real contributions does not cancel due to the different group theory factors
and  it brings out the Sudakov logarithm. It is known as the  Block-Nordsieck violation in 
electroweak processes. As a result, the PDF for nonsinglets contain the Sudakov logarithm
and does not satisfy the DGLAP equation, but a different evolution equation.
Furthermore, the factorization in weak interaction may be violated due to the Block-Nordsieck violation
Glauber exchange between spectator partons may violate factorization when the weak charges of the final
states are specified~\cite {Baumgart:2018ntv}. The possible breakdown of the factorization may 
start at order $\alpha^4$ of the magnitude $\sim \alpha^4 \ln^4 (M^2 /Q^2)$, 
and it is due to the fact that the group-theory factors for the exchange of two Glauber gauge bosons 
in different configurations across the unitarity cuts
are different and the overall effects do not cancel. This should be considered seriously in 
ascertaining the factorization in electroweak interaction, but it is beyond the scope of this paper, and 
has not been considered here. 

The nonsinglet beam functions contribute to the 2-jettiness, as well
as the singlet contributions. Due to the Sudakov logarithm and the existence of the
rapidity divergence, they obey coupled RG equations, and the evolution is distinct from that of the singlet
beam functions. It is also interesting to note that the additional contribution to the nonsinglet contributions
is proportional to $C_A$, whether it is the beam function (or the PDF) for the gauge bosons or for the lepton. 
And the matching coefficients which relate the PDF and the 
beam functions for the singlets and the nonsinglets are proportional to each other, both for the leptons and the 
gauge bosons. 
The nonsinglet soft functions get additional contributions, which are also proportional to $C_A$, and
they also satisfy the coupled RG equations.

We have considered the singlet and the nonsinglet contributions  to the 2-jettiness in both channels
$\ell_e \overline{\ell}_e \rightarrow \ell_{\mu} \overline{\ell}_{\mu}$ and $W W\rightarrow 
\ell_{\mu} \overline{\ell}_{\mu}$. The nonsinglet contributions  in $W W\rightarrow 
\ell_{\mu} \overline{\ell}_{\mu}$ and
$\ell_e \overline{\ell}_e \rightarrow \ell_{\mu} \overline{\ell}_{\mu}$ are suppressed due to
the additional evolution with respect to the rapidity scale, and they are about 50 \% of the singlet contributions
in size.  In $\ell_e \overline{\ell}_e \rightarrow \ell_{\mu} \overline{\ell}_{\mu}$, the nonsinglet contributions
alone are negative. But the physical jettiness is the total sum with the singlet contributions, and it is positive. 
In summary, though the nonsinglet contributions are suppressed, the nonsinglet 
contribution plays a significant role in the 2-jettiness numerically, and possibly in other jet shape observables.

In conclusion, we have established a consistent factorization for the 2-jettiness including the nonsinglet 
contributions, and their contributions are numerically significant. This result is in contrast to QCD,
in which there are only singlet contributions.
If we have more information on the beam function (or PDF) for the weak gauge bosons and the electrons, 
the prediction can be more precise.
 
We admit that the 2-jettiness analyzed here is far from realistic experimental comparison because we confine the
gauge group to SU(2). But the main focus here is to present clearly the difference between QCD and the 
weak interaction. And it turns out that the nonsinglet contributions in electroweak high-energy scattering
is appreciable, in comparison to QCD. Of course, we have to choose the electroweak gauge group 
SU(2)$\times$U(1), to be realistic. The phenomenology with the electroweak gauge group will be studied 
in the near future.

\appendix

\section{NLO calculation of the gauge-boson beam function and PDF\label{nlobf}}
 
\subsection{Beam functions for the gauge bosons} 
\begin{figure}[b] 
\begin{center}
\includegraphics[height=7cm]{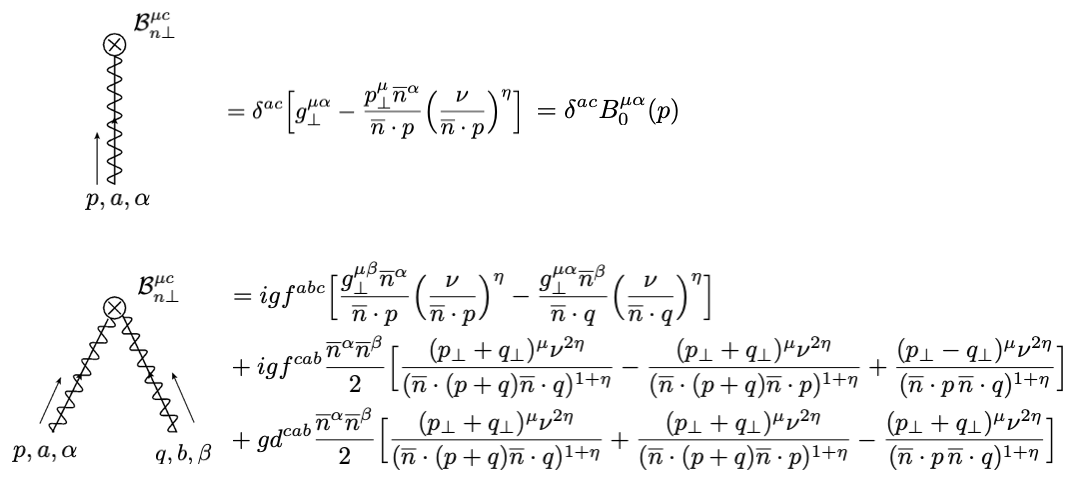} 
\end{center} \vspace{-0.6cm}
\caption{\label{wfeynman}\baselineskip 3.0ex Feynman rules for $\mathcal{B}_{n\perp}^{\mu c}$ with 
one and two gauge bosons.}
\end{figure}

\begin{figure}[t] 
\begin{center}
\includegraphics[height=7cm]{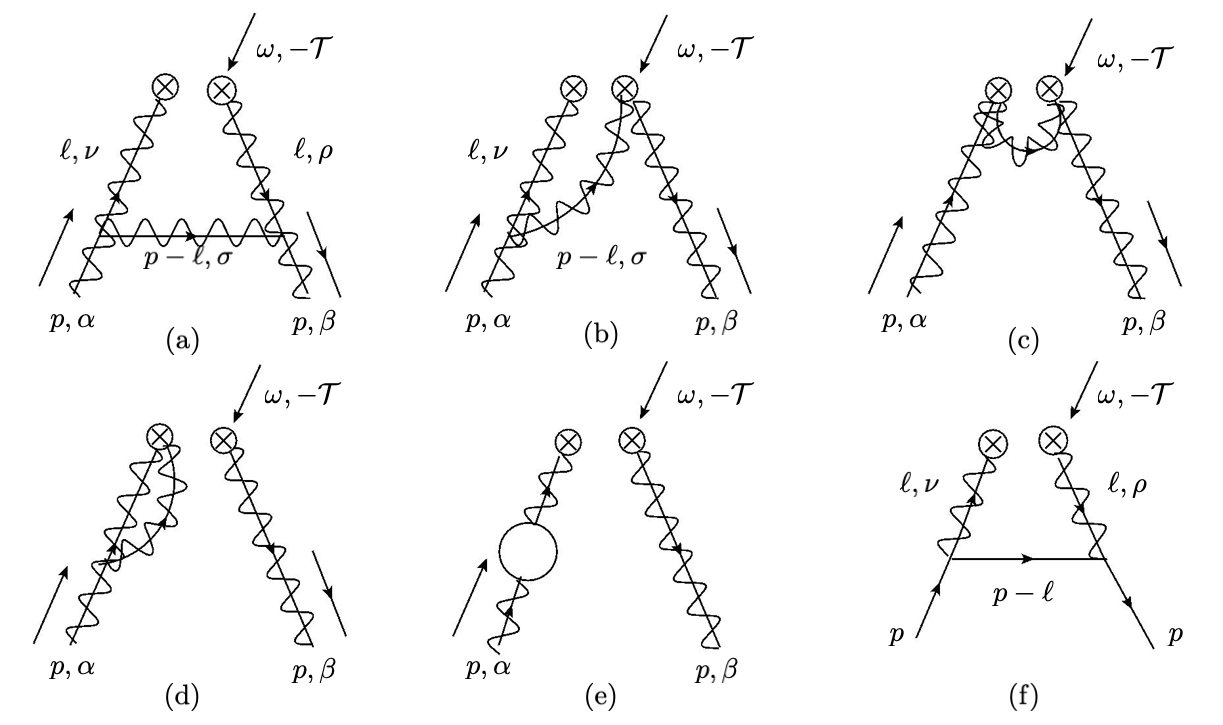} 
\end{center} \vspace{-0.6cm}
\caption{\label{wbeam}\baselineskip 3.0ex Feynman diagrams for the beam functions and the PDFs. The wavy
lines with solid lines denote collinear gauge bosons, and the solid lines are leptons. The mirror images of (b),
(d) and (e) are omitted. For the beam functions, the virtuality $t$ and the longitudinal momentum fraction $z$
are measured, but only  $z$ is measured for the PDFs. }
\end{figure}

The Feynman
rules for the gauge boson $\mathcal{B}_{n\perp}^{\mu c}$ get complicated due to the insertion of the 
rapidity regulator. The Feynman rules are shown in fig.~\ref{wfeynman}.
At NLO, the Feynman diagrams, contributing to the beam function, are presented in fig.~\ref{wbeam}.  
Figure~\ref{wbeam} (f) represents the mixing between the fermion and the gauge boson, but it does 
not possess any UV divergence, hence the beam function does not have any mixing.  In obtaining the final
result, we take the limit  of small mass $M$,  and keep only the logarithmic terms in $M$.
We also refer to fig.~\ref{wfeynman} for the Feynman diagrams of the PDFs with the caveat that $t$ is 
not measured for the PDFs.

Let us first consider the singlet matrix elements which are proportional to $(G^0)^{ab}$. Only the group theory
factors differ in the nonsinglet matrix elements. 
The naive contribution from  fig.~\ref{wbeam}(a) is written as
\begin{align}
&\tilde{M}_{(a)} (G^0)^{ab}= -\mms \int\dd{\ell} \eps_{\alpha}^a \eps_{\beta}^{b*} 
\frac{1}{(\ell^2 -M^2)^2}  V_{3\beta\rho\sigma}^{bde} (-p,\ell, p-\ell) V_{3\alpha\sigma\nu}^{aec} 
(p, \ell, -\-\ell)  \\
&\times B_0^{\mu\rho} (-\ell) B_{0\mu\nu} (\ell) (2\pi) \delta \Bigl( (p-\ell)^2 -M^2\Bigr) \delta 
\Bigl(\frac{t}{\omega} - (p^+ -\ell^+) \Bigr) \delta (\ell^- -\omega) \theta (\omega) \theta (t)
\theta(p^- -\ell^-), \nonumber 
\end{align}
where we average over the possible $(D-2)$ polarization, but not over the weak charges because we consider
the beam function with a fixed weak charge.
\begin{equation}
 \frac{1}{D-2} \sum_{\mathrm{pol.}} \eps_{\alpha}  \eps_{\beta}^{*} 
= -\frac{g_{\perp \alpha\beta}}{D-2}.
\end{equation}
And $V_{3\mu\nu\rho}^{abc} (p_1, p_2, p_3)$ is the three-boson vertex, which is given as
\begin{equation}
V_{3\mu\nu\rho}^{abc} (p_1, p_2, p_3) = gf^{abc} \Bigl[ g^{\mu\nu} (p_1 -p_2)^{\rho} +g^{\nu \beta} 
(p_2 -p_3)^{\mu} + g^{\rho\mu} (p_3 - p_1)^{\nu}\Bigr].
\end{equation}

Since $\tilde{M}_{(a)}$ is finite and we can put $\eps=\eta=0$ and perform the integration. 
After some algebra,  $\tilde{M}_{(a)}$ is given as
\begin{align} \label{tildema}
\tilde{M}_{(a)} &= -\frac{\alpha C_A}{4\pi}   \frac{1}{x(t-x^2 M^2)^2} 
\Bigl[ 2t (-2 +2x -3x^2 +2x^3) + M^2 x
(4+7x^2 +2x^3)\Bigr] \nonumber \\
&\longrightarrow \frac{\alpha C_A}{2\pi}  
\Bigl[ \Bigl( 2\frac{1-x}{x} + 2x (1-x) + \frac{3}{2} \frac{x(1-x)}{1-x+x^2}
\Bigr) \delta (t) \nonumber \\
&+ 2\Bigl( \frac{1-x}{x} + x(1-x) +\frac{x}{2}\Bigr) \Bigl( \delta (t) \ln \frac{(1-x)\mu^2}{x(1-x+x^2)M^2} 
+\frac{1}{\mu^2} \mathcal{L}_0\Bigl( \frac{t}{\mu^2}\Bigr) \Bigr) \Bigr].
\end{align}
In the small mass limit, it is regarded as the distribution in $t$. The coefficient of $\delta (t)$ is 
obtained by integrating $\tilde{M}_{(a)}$ over $t$. The remainder is determined by the fact that 
$\tilde{M}_{(a)}$ is independent of $\mu$.
The zero-bin contribution is suppressed, hence neglected at leading order in SCET. The functions 
$\mathcal{L}_n (x)$ are defined as
\begin{equation}\label{ldef}
\mathcal{L}_n (x) \equiv \Bigl[ \frac{\theta (x) \ln^n x}{x}\Bigr]_+ = \lim_{\beta\rightarrow 0} 
\Bigl[ \frac{\theta (x-\beta) \ln^n x}{x} +\delta (x-\beta) \frac{\ln^{n+1} \beta}{n+1}\Bigr].
\end{equation}

The naive contribution $\tilde{M}_{(b)}$ from fig.~\ref{wbeam}(b) is given as
\begin{align} \label{tildemb}
&\tilde{M}_{(b)} (G^0)^{ab}= \mms \int \dd{\ell}\frac{g_{\perp \alpha\beta}}{D-2} 
B_{1,bcd}^{\mu\rho\beta} (p-\ell,-p)
\delta (\ell^- -\omega) \delta (t+ \omega(p^+ -\ell^+)) \nonumber \\
&\times (2\pi)\delta ((p-\ell)^2 -M^2) V_{3\alpha\rho\nu}^{adc} (p, \ell-p,-\ell) \frac{1}{\ell^2 -M^2}
B_0^{\mu\nu} (\ell) \theta (\omega) \theta(p^- -\ell^-) \theta(t)  \nonumber \\
 &=\frac{\alpha C_A}{4\pi} (G^0)^{ab}\dms \int d\ell^+ d\ell^- d\bm{\ell}_{\perp}^2 
(\bm{\ell}_{\perp}^2)^{-\eps} \frac{\ell^- +p^-}{\ell^- -p^-} \frac{1}{\ell^2 -M^2} \Bigl( 
\frac{\nu^2}{\ell^- (p^- -\ell^-)}\Bigr)^{\eta} \nonumber 
\end{align}
\begin{align}
&\times \delta (\ell^- -\omega) \delta(\ell^+ -(p^+ -b^+)) \delta \Bigl( \bm{\ell}_{\perp}^2 -
\Bigl( \frac{(1-x)t}{x} -M^2\Bigr) \Bigr) \theta (\omega) \theta (p^- -\omega) \nonumber \\
&=\frac{\alpha C_A}{4\pi} (G^0)^{ab}\dms \frac{1+x}{1-x} \Bigl( \frac{(1-x)t}{x} -M^2\Bigr)^{-\eps} 
\frac{x}{t-x^2 M^2} \Bigl( \frac{\nu^2}{x(1-x)(p^-)^2}\Bigr)^{\eta} \theta (t).
\end{align}
In computing the group theory factor, we have used the fact that $f^{adc} f^{bcd} = - \delta^{ab}/N$.

Since it does not possess any divergence, we can put $\eps=\eta =0$, and it is written as
\begin{equation} \label{naimb}
\tilde{M}_{(b)} = \frac{\alpha C_A}{4\pi}  \frac{x(1+x)}{1-x} \frac{1}{t-x^2 M^2} 
\theta\Bigl( \frac{(1-x)t}{x} -M^2
\Bigr) \theta (x) \theta (1-x).
\end{equation}
In the limit of small $M$, it can be written as a double distribution in terms of $t$ and $x$.
In order to extract the coefficient for $\delta (t)$, we first look at the integral
\begin{equation} \label{mbtint}
\frac{x(1+x)}{1-x} \int_{xM^2/(1-x)}^{\mu^2} \frac{dt}{t-x^2 M^2} =\frac{x(1+x)}{1-x}  
\ln \frac{(1-x)(\mu^2 -M^2 x^2)}{x(1-x+x^2)M^2},
\end{equation}
where $\mu$ is chosen to be an arbitrary scale. The right-hand side of eq.~\eqref{mbtint} can be
expressed in terms of the distributions by considering the dependence on $x$ as
\begin{align} \label{xdist}
\frac{x(1+x)}{1-x}  
\ln \frac{(1-x)(\mu^2 -M^2 x^2)}{x(1-x+x^2)M^2} &= A\delta (1-x) + x(1+x) \mathcal{L}_0 (1-x)
\ln \frac{\mu^2}{x(1-x+x^2) M^2} \nonumber \\
&+x (1+x) \mathcal{L}_1 (1-x),
\end{align}
where $A$ is a constant to be determined. Integrating the left-hand side of eq.~\eqref{xdist} yields
\begin{equation}
\int_0^{k} dx \frac{x(1+x)}{1-x}  
\ln \frac{(1-x)(\mu^2 -M^2 x^2)}{x(1-x+x^2)M^2} = -\frac{9}{2} +\frac{5\sqrt{3}}{6} +\frac{4}{9}\pi^2
-\frac{5}{2}\ln \frac{\mu^2}{M^2} + \ln^2 \frac{\mu^2}{M^2},
\end{equation}
with $k=\mu^2/(\mu^2 +M^2)$, and integrating the right-hand side yields
\begin{align}
&\int_0^1 dx \Bigl[ A\delta (1-x) + x(1+x) \mathcal{L}_0 (1-x) \ln \frac{\mu^2}{x(1-x+x^2) M^2}
+x (1+x) \mathcal{L}_1 (1-x) \Bigr] \nonumber \\
&= A -\frac{9}{2} +\frac{5\sqrt{3}}{6} +\frac{4}{9}\pi^2
-\frac{5}{2}\ln \frac{\mu^2}{M^2}. 
\end{align}
By comparing two quantities, we obtain $A = \ln^2 (\mu^2/M^2)$.

As a result, we can write $\tilde{M}_{(b)}$ as
\begin{align}
&\tilde{M}_{(b)} =\frac{\alpha C_A}{4\pi} \Bigl\{ \delta (t) \Bigl[ \delta (1-x) 
\ln^2 \frac{\mu^2}{M^2} + x(1+x) \mathcal{L}_0 (1-x)
\ln \frac{\mu^2}{x(1-x+x^2) M^2} \nonumber \\
& + x(1+x) \mathcal{L}_1 (1-x)\Bigr]+ f_b (x,\mu) \frac{1}{\mu^2}\mathcal{L}_0 \Bigl(\frac{t}{\mu^2}\Bigr) 
+ g_b (x,\mu) \frac{1}{\mu^2}\mathcal{L}_1 \Bigl(\frac{t}{\mu^2}\Bigr) \Bigr\}.
\end{align}
The unknown functions $f_b (x,\mu)$ and $g_b (x,\mu)$ are determined by the requirement that 
$\tilde{M}_{(b)}$ is independent of $\mu^2$, that is, $d\tilde{M}_{(b)}/d\ln \mu^2 =0$.
They are given as
\begin{equation}
f_b (x, \mu) = 2 \delta (1-x) \ln \frac{\mu^2}{M^2} + x(1+x) \mathcal{L}_0 (1-x), \ \ 
g_b (x) = 2\delta (1-x).
\end{equation}
The final result for $\tilde{M}_{(b)}$ is written as
\begin{align}
\tilde{M}_{(b)} &= \frac{\alpha C_A}{4\pi}  \Bigl\{ \delta (t) 
\Bigl[ \delta (1-x) \ln^2 \frac{\mu^2}{M^2} 
+ x(1+x) \mathcal{L}_0 (1-x) \ln \frac{\mu^2}{x(1-x+x^2) M^2}\nonumber\\
& +x (1+x) \mathcal{L}_1 (1-x) \Bigr]  +\Bigl( 2\ln \frac{\mu^2}{M^2} 
\delta (1-x) + x(1+x) \mathcal{L}_0 (1-x)\Bigr) 
\frac{1}{\mu^2}\mathcal{L}_0 \Bigl(\frac{t}{\mu^2}\Bigr)  \nonumber \\
&+2\delta (1-x) 
\frac{1}{\mu^2}\mathcal{L}_1 \Bigl(\frac{t}{\mu^2}\Bigr) \Bigr\}. 
\end{align}
 
From eq.~\eqref{tildemb}, the zero-bin contribution $M_{(b)}^{\varnothing}$ can show up 
when $p-\ell$ or $\ell$
become soft. However, the contribution for the soft $\ell$ is suppressed and only the case with soft $p-\ell$ 
contributes. It is given, with $b^+ = t/\omega$, as
\begin{align}
M_{(b)}^{\varnothing}   
&= \frac{\alpha C_A}{4\pi}  \dms \delta (1-x) \int_{M^2/b^+}^{\infty} d\ell^- 
\frac{2}{b^+ p^- \ell^-} (\ell^- b^+ -M^2)^{-\eps} \Bigl(\frac{\nu}{\ell^-}\Bigr)^{\eta}  \nonumber \\
&= \frac{\alpha C_A}{2\pi}  \delta (1-x) 
\Bigl( \frac{\mu^2 e^{\gamma_{\mathrm{E}}}}{M^2}\Bigr)^{\eps}
\frac{\Gamma(\eps+\eta)}{\Gamma(1+\eta)} \Bigl(\frac{\nu^3}{\omega M^2}\Bigr)^{\eta} 
\frac{\nu^{-2\eta}}{t^{1-\eta}} \nonumber \\
&= \frac{\alpha C_A}{2\pi}  \delta (1-x) \Bigl\{ \delta (t)\Bigl[ 
\Bigl(\frac{1}{\eta} + \ln \frac{\nu}{p^-}\Bigr) 
\Bigl( \frac{1}{\eps} +\ln \frac{\mu^2}{M^2}\Bigr)  -\frac{1}{\eps^2} + \frac{1}{2} \ln^2 \frac{\mu^2}{M^2} 
+\frac{\pi^2}{12}\Bigr] \nonumber \\
&+ \Bigl( \frac{1}{\eps} +\ln \frac{\mu^2}{M^2}\Bigr) \frac{1}{\mu^2}\mathcal{L}_0 \Bigl(\frac{t}{\mu^2}\Bigr)
\Bigr\}.
\end{align}

The naive contribution $\tilde{M}_{(d)}$ is given as
\begin{align}
&\tilde{M}_{(d)} (G^0)^{ab}= -\mms \frac{g_{\perp\alpha\beta}}{2(D-2)}   \int \dd{\ell} B_0^{\mu\beta} (-p) 
B_{1,\mu\sigma\rho}^{bcd} (\ell, p-\ell) V_{3adc}^{\alpha\rho\sigma} (p,\ell -p,-\ell) \nonumber \\
&\times \frac{1}{\ell^2 -M^2} \frac{1}{(p-\ell)^2 -M^2} \delta (p^- -\omega) \delta (t/\omega) \nonumber \\
&= ig^2 \frac{(G^0)^{ab}C_A}{2} \mms \delta (1-x) \delta (t)  \int\dd{\ell} 
\frac{1}{\ell^2 -M^2} \frac{1}{(p-\ell)^2 -M^2} \nonumber \\
&\times \Bigl[ \frac{\ell^- +p^-}{p^- -\ell^-}\Bigl( \frac{\nu}{p^- -\ell^-}\Bigr)^{\eta}
+\frac{2p^- -\ell^-}{\ell^-} \Bigl(\frac{\nu}{\ell^-}\Bigr)^{\eta}\Bigr] \nonumber \\
&=ig^2  (G^0)^{ab}C_A  \mms \delta (1-x) \delta (t)  \int\dd{\ell} 
\frac{1}{\ell^2 -M^2} \frac{1}{(p-\ell)^2 -M^2} \frac{\ell^- +p^-}{p^- -\ell^-}
\Bigl( \frac{\nu}{p^- -\ell^-}\Bigr)^{\eta}\nonumber 
\end{align}
\begin{align}
&= -\frac{\alpha C_A}{4\pi} (G^0)^{ab} \delta (1-x) \delta (t) \Gamma(\eps) 
\Bigl(\frac{\mu^2}{M^2}\Bigr)^{\eps} \Bigl(\frac{\nu}{p^-}\Bigr)^{\eta}\int_0^1 dy  
\frac{1+y}{(1-y)^{1+\eta}} (1-y+y^2)^{-\eps}. 
\end{align}
The zero-bin contribution $M_{(d)}^{\varnothing}$ comes from the region in which 
$p^{\mu} -\ell^{\mu}$ becomes
soft, while the contribution from the region in which $\ell^{\mu}$ becomes soft is suppressed. It is given as
\begin{align}
M_{(d)}^{\varnothing} &= \frac{\alpha C_A}{4\pi}   \delta (1-x) \delta (t)  
e^{\gamma_{\mathrm{E}}\eps}  \Bigl(\frac{\mu^2}{M^2}\Bigr)^{\eps} \Bigl(\frac{\nu}{M}\Bigr)^{\eta}
\frac{\Gamma (\eps+\eta/2)}{\Gamma(1+\eta/2)} \int_0^{\infty} y^{-1+\eta}.
\end{align}
The net contribution $M_{(d)}=\tilde{M}_{(d)} - M_{(d)}^{\varnothing}$ is computed and the result is 
expanded near $\eta =0$ first, and then near $\eps=0$, and the result is given as
\begin{equation}
M_{(d)}  = \frac{\alpha C_A}{4\pi}   \delta (1-x) \delta (t)
\Bigl[ \Bigl( \frac{1}{\eps} +\ln \frac{\mu^2}{M^2}\Bigr) \Bigl( \frac{2}{\eta} + 2\ln \frac{\nu}{p^-} +1\Bigr)
+2 -\frac{\pi}{\sqrt{3}} -\frac{\pi^2}{9}\Bigr].
\end{equation}

\begin{figure}[b] 
\begin{center}
\includegraphics[height=4.5cm]{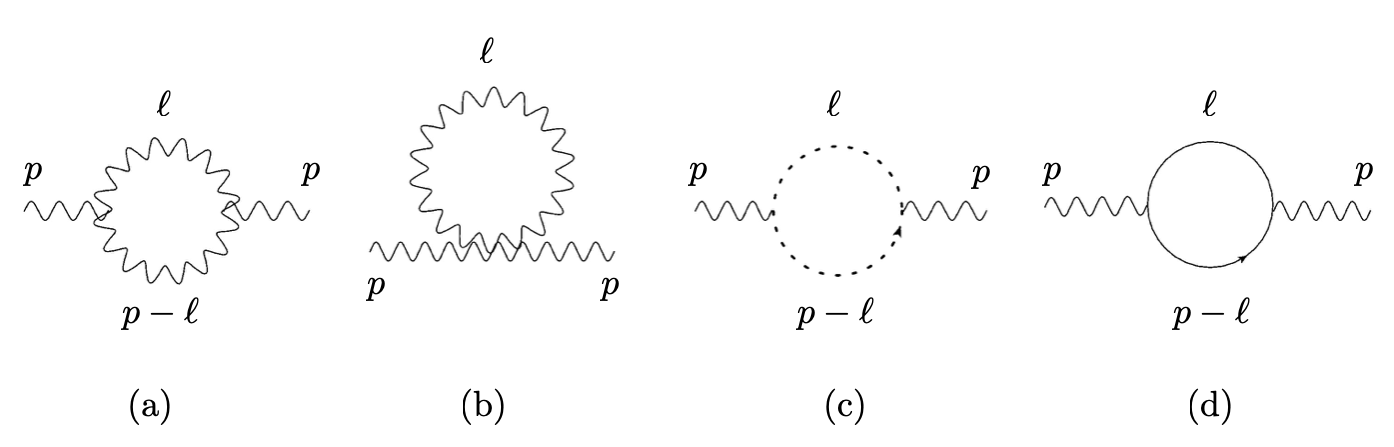} 
\end{center} \vspace{-0.6cm}
\caption{\label{wself1}\baselineskip 3.0ex Feynman diagrams for the self-energy of the gauge bosons
from the loops of (a, b) the gauge bosons, (c)  the ghost loops, and (d)  the fermions.}
\end{figure}

\begin{figure}[t] 
\begin{center}
\includegraphics[height=4.5cm]{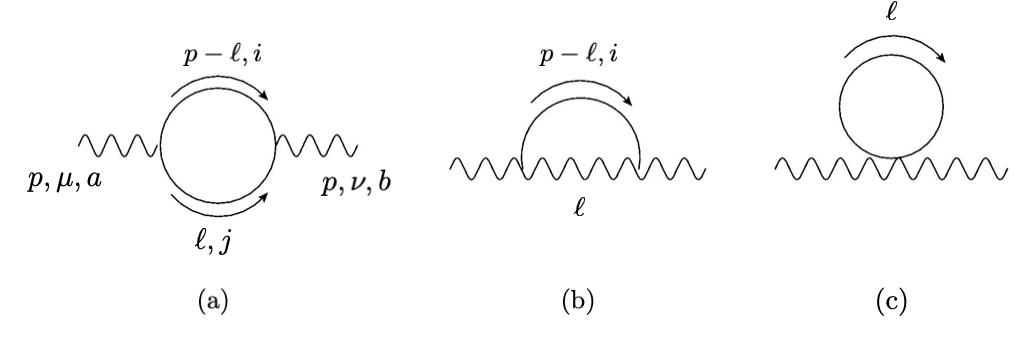} 
\end{center} \vspace{-0.6cm}
\caption{\label{scalarmass}\baselineskip 3.0ex Feynman diagrams for the self-energy of the gauge bosons
from   the scalar particles.  (a) contains a divergence, (b) is finite, and (c) does not contribute to the 
field-strength renormalization.}
\end{figure}

In fig.~\ref{wbeam}(e), it involves the self-energy of the gauge boson and the blob contains the loops
of the gauge bosons, the ghosts, the fermions and the scalar particles. The Feynman diagrams 
are shown in figs.~\ref{wself1} and \ref{scalarmass}.
Contrary to QCD, we compute the self-energy corrections with the gauge boson mass $M$ with the left-handed
fermions only. And the 
additional contributions come from the interacton of the complex scalar multiplets.

The self-energy of the gauge boson can be written as
\begin{equation}
\Pi (p^2) \delta^{ab} \Bigl( g^{\mu\nu} - \frac{p^{\mu} p^{\nu}}{p^2}\Bigr), 
\end{equation}
and the field-strength renormalization $Z$ is given as
\begin{equation}
Z = \Bigl( 1-\frac{d\Pi (p^2)}{dp^2}\Big|_{p^2 =M^2}\Bigr)^{-1}.
\end{equation}
The field-strength renormalization $Z_W$ from the gauge bosons and the ghost particles, common to
all the SU($N$) gauge interactions, comes from fig.~\ref{wself1} (a) to (c). At order $\alpha$, it is given as
\begin{equation}
Z_W^{(1)} = \frac{\alpha C_A}{4\pi} \Bigl[ \frac{5}{3} \Bigl( \frac{1}{\eps} +\ln \frac{\mu^2}{M^2}\Bigr)
-\frac{5}{9} +\frac{\pi}{\sqrt{3}}\Bigr].
\end{equation}
The contribution $Z_f$ from the fermion loop in fig.~\ref{wself1} (d) at order $\alpha$ is given by
\begin{equation}
Z_f ^{(1)} = \frac{\alpha}{4\pi} \frac{1}{2} n_f T_F \Bigl[ -\frac{4}{3} \Bigl( \frac{1}{\eps} 
+\ln \frac{\mu^2}{M^2}\Bigr) -\frac{8}{9}\Bigr].
\end{equation}
The factor 1/2 is to remind that only the left-handed fermions contribute.

The remaining contribution comes from the scalar particles. 
For the SU(2) gauge group, we follow the standard scalar fields in the Standard Model. However, for
general SU($N$) gauge theories, the symmetry breaking pattern may depend on the model. We just assume
that  all the gauge bosons have a common mass $M$, as in SU(2). The Feynman diagrams 
from the scalar particles are shown in fig.~\ref{scalarmass}. Fig.~\ref{scalarmass}(a) contains an UV divergence, and its contribution is denoted as $Z_{sa}$ below.
Fig.~\ref{scalarmass}(b) depends on the structure of the complex scalar multiplets, to yield the same
masses to all the gauge bosons in the SU($N$) gauge theory. However, since it is finite and does not contribute to
the field-strength renormalization, we neglect this. Note that this contribution cancels in obtaining
the matching coefficients in eq.~\eqref{iww}. And fig.~\ref{scalarmass}(c) vanishes.

The field-strength renormalization $Z$ at order $\alpha$ is given as
\begin{align}
Z^{(1)} &= Z_W^{(1)}  + Z_f^{(1)} + Z_{sa}^{(1)}   \\
&= \frac{\alpha}{4\pi} \Bigl[ (\beta_0 -2C_A) \Bigl( \frac{1}{\eps} +\ln \frac{\mu^2}{M^2}\Bigr) +C_A
(-\frac{5}{9} +\frac{\pi}{\sqrt{3}}\Bigr) - \frac{4}{9} n_f T_F  + n_s T_F \Bigl( -\frac{17}{9} +\frac{\pi}{\sqrt{3}}   \Bigr) \Bigr], \nonumber
\end{align}
where $\beta_0 = 11C_A/3 - 2 n_f T_F /3 -n_s T_F/3$ is the beta function at leading order.

Adding all the contributions, the singlet beam function $B_{Ws}^{(1)}$ for the gauge boson at NLO is written 
as
\begin{align}
&B_{Ws}^{(1)} (t,x,\mu)  = M_{(a)} + 2(M_{(b)} -M_{(b)}^{\varnothing} +M_{(d)} - M_{(d)}^{\varnothing})
+Z^{(1)}\delta (1-x) \delta (t) \nonumber \\
&=\frac{\alpha}{2\pi} \Bigl\{ \delta (1-x) \Bigl[ C_A \Bigl( \frac{2}{\eps^2} \delta (t) -
\frac{2}{\eps} \frac{1}{\mu^2} \mathcal{L}_0 \bigl( \frac{t}{\mu^2}\bigr) \Bigr) 
+\frac{\beta_0}{2} \frac{1}{\eps}\Bigr] \nonumber  \\
&+\delta (t) \Bigl( \frac{\beta_0}{2} \delta (1-x) +C_A P_{WW} (x)\Bigr) \ln \frac{\mu^2}{M^2}
+C_A \delta (t)\Bigl( P_{WW} (x) \ln \frac{1-x}{x} -\frac{\pi^2}{6} \delta (1-x)\Bigr)
 \nonumber \\
&+ C_A \Bigl[ \delta (1-x) \frac{2}{\mu^2} \mathcal{L}_1 \bigl( \frac{t}{\mu^2}\bigr) + P_{WW} (x)
\frac{1}{\mu^2} \mathcal{L}_0 \bigl( \frac{t}{\mu^2}\bigr)\Bigr] \nonumber \\
 &+C_A\delta (t) \Bigl[ \delta (1-x) \Bigl( \frac{31}{18} -\frac{\pi^2}{9} -\frac{\pi}{2\sqrt{3}}\Bigr)
-\Bigl(\frac{2(1-x)}{x} + 2x(1-x) +\frac{3}{2}\frac{x(1-x)}{1-x+x^2}\Bigr) \nonumber \\
&-P_{WW} (x) \ln (1-x+x^2)\Bigr] + \delta (1-x) \delta (t) \Bigl(-\frac{2}{9} n_f T_F + \frac{1}{2} n_s T_F
\Bigl(-\frac{17}{9} +\frac{\pi}{\sqrt{3}}\Bigr)\Bigr\}.
\end{align}

For the nonsinglet beam functions, only the group theory factors are different. Extracting the nonsinglet
matrix elements, proportional to $d^{abc}$,  the matrix elements $\tilde{M}_{(a)}$, 
$\tilde{M}_{(b)}$ and $M_{(b)}^{\varnothing}$ in the singlet calculation are replaced by $\tilde{M}_{(a)}/2$, 
$\tilde{M}_{(b)}/2$ and $M_{(b)}^{\varnothing}/2$ respectively, while the remaining matrix elements 
are the same. Therefore, the nonsinglet beam functions at NLO are given as
\begin{align}
&B_{Wn}^{(1)} (t,x,\mu,\nu) = \frac{1}{2} \tilde{M}_{(a)} + \tilde{M}_{(b)} - M_{(b)}^{\varnothing} +
2(\tilde{M}_{(d)} - M_{(d)}^{\varnothing}) + Z^{(1)} \delta (1-x) \delta (t) \nonumber \\
& =B_{Ws}^{(1)} (t,x,\mu) -\frac{1}{2} \Bigl( \tilde{M}_{(a)} + 2(\tilde{M}_{(b)} - M_{(b)}^{\varnothing} )\Bigr)
\nonumber \\
& =B_{Ws}^{(1)} (t,x,\mu)  \nonumber \\
&+\frac{\alpha C_A}{4\pi}\Bigl\{ 2\delta (1-x) \Bigl[ \delta (t) \Bigl( \frac{1}{\eps} 
+ \ln \frac{\mu^2}{M^2}\Bigr) \Bigl( \frac{1}{\eta}
+\ln \frac{\nu}{p^-}\Bigr) -\frac{1}{\eps^2} \delta (t)
+ \frac{1}{\eps} \frac{1}{\mu^2} \mathcal{L}_0 \bigl( \frac{t}{\mu^2}\bigr)  -\frac{1}{\mu^2}\mathcal{L}_1 
\bigl(\frac{t}{\mu^2}\bigr)\Bigr] \nonumber \\
&-P_{WW} (x)\Bigl( \delta (t) \ln \frac{\mu^2}{M^2} +\frac{1}{\mu^2} \mathcal{L}_0 
\bigl( \frac{t}{\mu^2}\bigr)  \Bigr) +\delta (t) \Bigl( P_{WW}(x)  \ln \frac{x(1-x+x^2)}{1-x}+\frac{\pi^2}{6}
\delta(1-x)   \nonumber \\
&+   2\frac{1-x}{x} + 2x(1-x) +\frac{3}{2} \frac{x(1-x)}{1-x+x^2}\Bigr) \Bigr\}.
\end{align}

\subsection{PDF for the gauge bosons}

We can compute the PDF either by computing the matrix elements
in eq.~\eqref{wpdf2},  or by integrating the results  of the beam functions
with respect to the variable $t$. This can be seen by comparing the definitions of the beam functions in 
eq.~\eqref{wbeam2} and those of the PDFs in eq.~\eqref{wpdf2}. We adopt the second approach here.

We first consider the matrix elements for the singlet PDF.  
The naive contribution from 
fig.~\ref{wbeam}(a) can be obtained from eq.~\eqref{tildema} by integrating with respect to $t$,
 and it is given as
\begin{align}
\tilde{M}_{(a)} &= \frac{\alpha C_A}{2\pi}  \Bigl[ 2\Bigl(\frac{1-x}{x} + \frac{3x}{2} -x^2\Bigr)  
\Bigl(\frac{1}{\eps} +\ln \frac{ \mu^2}{ (1-x+x^2)M^2} \Bigr)  \nonumber \\
&+\frac{2(1-x)}{x} +2x(1-x) +\frac{3}{2} \frac{x(1-x)}{1-x+x^2} \Bigr].
\end{align}

Integrating eq.~\eqref{tildemb} with respect to $t$ yields $\tilde{M}_b$ for the PDF. However, the correct
rapidity divergence is obtained only after the zero-bin subtraction is performed. We present the 
true collinear contribution $M_b = \tilde{M}_b -M^{\varnothing}_b$ as 
\begin{align}
M_{(b)} &= \frac{\alpha C_A}{4\pi}  \Bigl[ -2\Bigl(\frac{1}{\eta} +  \ln \frac{\nu}{p^-}\Bigr)
\Bigl( \frac{1}{\eps} +\ln \frac{\mu^2}{M^2}\Bigr)  + x(1+x) \mathcal{L}_0 (1-x) 
\Bigl( \frac{1}{\eps} +\ln \frac{\mu^2}{M^2 (1-x+x^2)}\Bigr) \Bigr].
\end{align}
In a similar approach to computing $M_{(b)}$, $M_{(d)}$ can be obtained as
\begin{equation}
M_{(d)} = \frac{\alpha C_A}{4\pi}  \delta (1-x) \Bigl[ \Bigl(\frac{2}{\eta} + 2\ln \frac{\nu}{p^-}
+1\Bigr) \Bigl( \frac{1}{\eps} +\ln \frac{\mu^2}{M^2}\Bigr)  + 2 -\frac{\pi}{\sqrt{3}} -\frac{\pi^2}{9}\Bigr].
\end{equation}

The wave function renormalization $M_{(e)}$ is given by
\begin{align}
Z^{(1)} &=\frac{\alpha}{4\pi} \Bigl[ (\beta_0 -2C_A) \Bigl( \frac{1}{\eps} +\ln \frac{\mu^2}{M^2}\Bigr) 
+C_A \Bigl( -\frac{5}{9}+\frac{\pi}{\sqrt{3}}\Bigr)-\frac{4}{9}  n_f T_F + n_s T_F \Bigl( -\frac{17}{9} + \frac{\pi}{\sqrt{3}}  \Bigr) \Bigr].
\end{align}

Combining all the matrix elements, the singlet PDF at NLO is given as
\begin{align}
& f_{Ws}^{(1)} (x,\mu, \nu) = \tilde{M}_a + 2(M_b + M_d) +Z^{(1)}\delta (1-x) \nonumber \\
&= \frac{\alpha}{2\pi}   \Bigl\{ C_A \Bigl(P_{WW} (x) +\frac{\beta_0}{2} \delta (1-x)\Bigr) \Bigl( \frac{1}{\eps}
+ \ln \frac{\mu^2}{M^2}\Bigr)  \nonumber \\
& -C_A \Bigl(P_{WW} (x) \ln (1-x+x^2) +2\frac{1-x}{x} +2x(1-x) +\frac{3}{2} \frac{x(1-x)}{1-x+x^2} 
\Bigr)   \nonumber \\
& +\delta (1-x) \Bigl[ C_A \Bigl( \frac{31}{18} -\frac{\pi}{2\sqrt{3}} -\frac{\pi^2}{9}\Bigr) 
- \frac{2}{9} n_f   T_F    + n_s T_F \Bigl( -\frac{17}{9} +\frac{\pi}{\sqrt{3}}  \Bigr) \Bigr\}. 
\end{align}

The matrix elements for the nonsinglet PDFs have different group theory factors, and the relation between the
singlets and the nonsinglets is the same as that in the beam functions. The nonsinglet PDFs at NLO are written as
\begin{align}
& f_{Wn}^{(1)} (x,\mu,\nu) = \frac{1}{2} M_a + \tilde{M}_{(b)} - M_{(b)}^{\varnothing} +
2(\tilde{M}_{(d)} - M_{(d)}^{\varnothing}) + Z^{(1)} \delta (1-x)   \nonumber \\
& =f_{Ws}^{(1)} (x,\mu) -\frac{1}{2} \Bigl( \tilde{M}_{(a)} + 2(\tilde{M}_{(b)} - M_{(b)}^{\varnothing} )\Bigr)
\nonumber \\
&=f_{Ws}^{(1)} (x,\mu)+\frac{\alpha}{2\pi}\frac{C_A}{2}   \Bigl[ \delta(1-x)  
\Bigl(\frac{2}{\eta} +2\ln \frac{\nu}{p^-}\Bigr) \Bigl( \frac{1}{\eps} +\ln \frac{\mu^2}{M^2}\Bigr)  
-  \Bigl(\frac{1}{\eps} +\ln \frac{\mu^2}{M^2}\Bigr)  P_{WW} (x) \nonumber \\
&+2\frac{1-x}{x} +2x (1-x) +\frac{3}{2} \frac{x(1-x)}{1-x+x^2}  +P_{WW} (x) \ln (1-x+x^2)\Bigr].
\end{align}

\section{Soft matrix elements at tree level\label{softzm}}
When there are singlets only, the soft matrix is given by
\begin{equation}
S^{(0)} (0, 0, 0, 0) =
\begin{pmatrix}
N(N^2-1)&0&0 \\
0&\frac{1}{2}N(N^2-1) &0 \\
0&0&\frac{(N-2)(N+2)(N^2-1)}{2N}
\end{pmatrix}.  
\end{equation}
When there are 2 nonsinglets, the soft matrices are given as
\begin{align}
S^{(0)} (1, 1, 0, 0) &=(N-2)
\begin{pmatrix}
2(N+2)&0&0\\
0&\frac{1}{2}(N+2)&0\\
0&0&\frac{(N+2)(N^2-12)}{2N^2}
\end{pmatrix}  (12), \nonumber\\
S^{(0)} (1, 0, 1, 0) &=(N-2)
\begin{pmatrix}
0&0&\frac{N+2}{N} \\
0&\frac{1}{4}(N+2)&\frac{1}{4}(N+2) \\
\frac{N+2}{N}&\frac{1}{4}(N+2)&\frac{(N+2)(N^2-12)}{4N^2}
\end{pmatrix} (13), \nonumber \end{align}
\begin{align}
S^{(0)} (1, 0, 0, 1) &=(N-2)
\begin{pmatrix}
0&0&\frac{N+2}{N} \\
0&\frac{1}{4}(N+2)&-\frac{1}{4}(N+2) \\
\frac{N+2}{N}&-\frac{1}{4}(N+2)&\frac{(N+2)(N^2-12)}{4N^2}
\end{pmatrix} (14), \nonumber  \\
S^{(0)} (0, 1, 1, 0) &=(N-2)
\begin{pmatrix}
0&0&\frac{N+2}{N} \\
0&\frac{1}{4}(N+2)&-\frac{1}{4}(N+2) \\
\frac{N+2}{N}&-\frac{1}{4}(N+2)&\frac{(N+2)(N^2-12)}{4N^2}
\end{pmatrix} (23) ,\nonumber \end{align}
\begin{align}
S^{(0)} (0, 1, 0, 1) &=(N-2)
\begin{pmatrix}
0&0&\frac{N+2}{N} \\
0&\frac{1}{4}(N+2)&\frac{1}{4}(N+2) \\
\frac{N+2}{N}&\frac{1}{4}(N+2)&\frac{(N+2)(N^2-12)}{4N^2}
\end{pmatrix} (24), \nonumber \\
S^{(0)} (0, 0, 1, 1) &=
\begin{pmatrix}
(N^2-1)&0&0 \\
0&-\frac{1}{2} &0 \\
0&0&-\frac{(N-2)(N+2)}{2N^2}
\end{pmatrix} (34).
\end{align}
Note that, with $N=2$, only the last term $S^{(0)} (0,0,1,1)$ survives, which correspond to 
the case with no nonsinglet beam functions and 2 jet functions.

For the case of 3 nonsinglets, they are given as
\begin{align}
S^{(0)} (1, 1, 1, 0) &= (N-2)
\begin{pmatrix}
0&\frac{N+2}{2N}\big((123)-(132)\big)&\frac{N^2-12}{2N}\frac{(123)+(132)}{N-2} \\
-\frac{N+2}{2N}\big((123)-(132)\big)&-\frac{(123)+(132)}{N-2}&\frac{N+2}{N^2}\big((123)-(132)\big) \\
\frac{N^2-12}{2N}\frac{(123)+(132)}{N-2}&-\frac{N+2}{N^2}\big((123)-(132)\big) &-\frac{2(N^2-10)}{N^2} \frac{(123)+(132)}{N-2}
\end{pmatrix}, \nonumber \\
S^{(0)} (1, 1, 0, 1) &= (N-2)
\begin{pmatrix}
0&\frac{N+2}{2N}\big((124)-(142)\big)&\frac{N^2-12}{2N}\frac{(124)+(142)}{N-2} \\
-\frac{N+2}{2N}\big((124)-(142)\big)&-\frac{(124)+(142)}{N-2}&\frac{N+2}{N^2}\big((124)-(142)\big) \\
\frac{N^2-12}{2N}\frac{(124)+(142)}{N-2}&-\frac{N+2}{N^2}\big((124)-(142)\big) &-\frac{2(N^2-10)}{N^2} \frac{(124)+(142)}{N-2}
\end{pmatrix}, \nonumber 
\end{align}
\begin{align}
S^{(0)} (1, 0, 1, 1) &= (N-2)
\begin{pmatrix}
0&0&\frac{N+2}{N} (143) \\
0&-\frac{1}{2} \frac{(134)+(143)}{N-2}&0 \\
\frac{N+2}{N} (134)& 0 &-\frac{N^2-12}{2N^2} \frac{(134)+(143)}{N-2}
\end{pmatrix}, \nonumber\\
S^{(0)} (0, 1, 1, 1) &= (N-2)
\begin{pmatrix}
0&0&\frac{N+2}{N} (243) \\
0&-\frac{1}{2} \frac{(234)+(243)}{N-2}&0 \\
\frac{N+2}{N} (234)& 0 &-\frac{N^2-12}{2N^2} \frac{(234)+(243)}{N-2}
\end{pmatrix}.
\end{align}
For SU(2), all the soft matrices with an odd number of nonsinglets vanish.

Finally, when there are 4 nonsinglets, they are given as
\begin{align}
S^{(0)} (1, 1, 1, 1) &= (N-2)
\begin{pmatrix}
\frac{2(N+2)}{N} (12)(34)&\frac{N+2}{2N} \big((1243)-(2143)\big)&S_{13} \\
-\frac{N+2}{2N} \big((1234)-(2134)\big)&S_{22}&S_{23}\\
S_{31}&S_{32}&S_{33}
\end{pmatrix}, \nonumber \end{align}\\
where $S_{ij}$ are given by
\begin{align}
S_{13}&=\frac{N^2-12}{2N^2}\frac{N\big((1243)+(2143)\big)-2 (12)(34)}{N-2}, \nonumber \\
S_{22}&=\frac{N}{4(N-2)} \Bigl((14)(23)+(13)(24) -2  (1324)-2(1423) -  (1234)-(1243)-(1342)-(1432)
\Bigr), \nonumber \\
S_{23}&=\frac{1}{4N^2 (N-2)}\ 
 \Bigl[ 16\Bigl((1342)-(1234)\Bigr)+N(N^2-8) \Bigl((13)(24)-(14)(23)\Bigr) \nonumber \\
&+N^2\Bigl( (1234)-(1243)-(1342)+(1432)\Bigr)\Bigr], \nonumber 
\end{align}
\begin{align}
S_{31}&=\frac{N^2-12}{2N^2(N-2)}\Bigl[N\Bigl((1234)+(2134)\Bigr)-2 (12)(34)\Bigr], \nonumber \\
S_{32}&=\frac{1}{4N^2 (N-2)}  \Bigl[ 16\Bigl((1243)-(1432)\Bigr)+N(N^2-8) \Bigl((13)(24)-(14)(23) \Bigr)
 \nonumber \\
&+N^2\Bigl((1234)-(1243)-(1342)+(1432)\Bigr)  \Bigr],\nonumber \\
S_{33}&=\frac{1}{4N^3 (N-2)}  \Bigl[ -64 (12)(34)+16 N\Bigl(2 (1234)+2 (1243)+2 (1342)+2 (1432))\nonumber \\
& +(1324)+(1423)\Bigr) +N^2 (N^2-16) \Bigl((14)(23)+(13) (24)\Bigr)\nonumber \\
&  -N^3\Bigl((1234)+(1243)+(1342)+(1432)-2(1423)-2(1324)\Bigr) \Bigl].
\end{align}

\acknowledgments{This work is supported by Basic Science Research Program through the National 
Research Foundation of Korea (NRF) funded by 
the Ministry of Education (Grant No. NRF-2019R1F1A1060396). T. K.  is supported by Basic Science 
Research Program through the National Research Foundation of Korea (NRF) funded by the Ministry of 
Science and ICT (Grants No. NRF-2021R1A2C1008906).}

\bibliographystyle{JHEP1}
\bibliography{jettiness1}
\end{document}